\documentclass{jfm}
\usepackage{graphicx}
\usepackage{epstopdf, epsfig}
\usepackage{subfigure}
\usepackage{amsmath}
\usepackage{amssymb}
\usepackage{amsfonts}
\usepackage{bm}
\usepackage{empheq}
\usepackage{latexsym}
\usepackage{cases}
\usepackage{multirow}

\shorttitle{Reflection of a jet K-H wave incident on a shock}
\shortauthor{M. Mancinelli et al.}

\title{Reflection and transmission of a Kelvin-Helmholtz wave incident on a shock in a jet}

\author{Matteo Mancinelli\aff{1}
  \corresp{\email{matteo.mancinelli@univ-poitiers.fr}},
  Eduardo Martini\aff{1}$^,$\aff{2},
  Vincent Jaunet\aff{1},
  Peter Jordan\aff{1},
  Aaron Towne\aff{3},
  \and Yves Gervais\aff{1}}

\affiliation{\aff{1}D\'{e}partement Fluides Thermique et Combustion, Institut Pprime - CNRS-Universit\'{e} de Poitiers-ENSMA, 11 Boulevard Marie et Pierre Curie, 86962 Chasseneuil-du-Poitou, France
\aff{2}Direction des Applications Militaires, CEA-Cesta, 15 Avenue des Sablières, 33114 Le Barp, France
\aff{3}Department of Mechanical Engineering, University of Michigan, 2350 Hayward Street, Ann Arbor, MI 48109, USA}

\begin{document}

\maketitle

\begin{abstract}
Screech tones in supersonic jets are underpinned by resonance between downstream-travelling Kelvin-Helmholtz waves and upstream-travelling acoustic waves. Specifically, recent work suggests that the relevant acoustic waves are guided within the jet and are described by a discrete mode of the linearised Navier-Stokes equations. However, the reflection mechanism that converts downstream-travelling waves into upstream-travelling waves, and vice versa, has not been thoroughly addressed, leading to missing physics within most resonance models. In this work we investigate the reflection and transmission of waves generated by the interaction between a Kelvin-Helmholtz wave and a normal shock in an under-expanded jet using a mode-matching approach. Both vortex-sheet and finite-thickness shear-layer models are explored, quantifying the impact of the shear layer in the reflection process. This approach could enable more quantitative predictions of resonance phenomena in jets and other fluid systems.
\end{abstract}

\begin{keywords}
Authors should not enter keywords on the manuscript, as these must be chosen by the author during the online submission process and will then be added during the typesetting process (see http://journals.cambridge.org/data/\linebreak[3]relatedlink/jfm-\linebreak[3]keywords.pdf for the full list)
\end{keywords}

\section{Introduction}
\label{sec:intro}
Shock-containing shear flows involve a rich variety of phenomena, and among these is shock-turbulence interaction (STI). In free shear layers, STI leads to an increase in turbulence levels and mixing downstream of the shock \citep{genin2010studies}. In wall-bounded flows, in addition to the enhancement of turbulence, STI may also be accompanied by boundary layer separation and the formation of a separation bubble \citep{delery1983experimental,dolling2001fifty,clemens2014low}. STI is also an important feature of supersonic combustion in scramjets \citep{yang1993applications}. In imperfectly expanded propulsive jets, STI underpins the generation of broad-band shock associated noise \citep{tanna1977experimental, tam1982shock} and screech \citep{powell1953mechanism, tam1986proposed, raman1999supersonic, edgington2019aeroacoustic}.

Linear theory has been widely used to study the interaction between disturbance fields and shocks. \citet{ribner1954convection} considered the interaction between a vorticity wave and a normal shock. The analysis was later extended to consider STI, where a homogeneous turbulence was modelled as a superposition of Fourier vorticity waves \citep{ribner1955shock}. \citet{moore1954unsteady} considered  interaction between sound waves and an oblique shock, and this work was extended by \citet{mahesh1995interaction} to study an isotropic field of acoustic disturbances interacting with a shock. Later, \cite{mahesh1997influence} considered the influence of entropy fluctuations on STI as well and \citet{buttsworth1996interaction} derived expressions for shock-induced vorticity, useful for the estimation of mixing enhancement. The foregoing studies were all based on solution of the Rankine-Hugoniot relations. The unsteady STI was converted into an equivalent steady-flow problem which did not consider the reflection process associated with the incident turbulent disturbance but only the transmission mechanism through the shock wave. A review of these studies and others has been compiled by \cite{andreopoulos2000shock}. More recently, \cite{kitamura2016rapid} used rapid distortion theory to study the interaction between homogeneous isotropic turbulence and a shock wave and \cite{chen2019shock} considered STI at high turbulence intensities. Similarly to the works reported above, the authors mainly focused on the turbulence amplification and modification of the turbulence length scales downstream of the shock.

The problem we consider is motivated by the sound generated by imperfectly expanded, supersonic jets, and, in particular, the phenomenon known as screech, a mechanistic explanation for which was first provided by \citet{powell1953mechanism}. The mechanism involves turbulent structures that are convected through the shock-cell structure; this STI results in the generation of upstream-travelling sound waves. According to Powell's phenomenological description, when the phases of the upstream-travelling sound waves and downstream-travelling turbulent structures are suitably matched, at the jet exit plane and at the STI locations, resonance may occur. The downstream-travelling turbulent structures considered important for screech are what are often referred to as coherent structures.

A large body of recent work has shown how coherent structures in turbulent jets, and the sound they produce, can be modelled using linear theory \citep{jordan2013wave,schmidt2017wavepackets,towne2018spectral,cavalieri2019wave,lesshafft2019resolvent,nogueira2019large,edgington2021waves}. As shown in these studies, downstream-travelling coherent structures are largely underpinned by Kelvin-Helmholtz (K-H) instability. As mentioned above, \cite{powell1953mechanism} assumed that the upstream-travelling waves responsible for the feedback mechanism in screech generation were free-stream acoustic waves. But this has been recently questioned. \citet{shen2002three} suggested that the upstream-travelling disturbance might comprise a family of guided jet modes, first discussed by \citet{tam1989three}. This hypothesis has been recently confirmed in studies by \citet{gojon2018oscillation} and \citet{edgington2018upstream}, and a simplified screech-tone prediction model based on this idea has been developed and validated by \citet{mancinelli2019screech}. In the simplest formulation of the screech-tone model, the spatial growth of the K-H mode is ignored, and a phase-matching criterion is sufficient to provide a reasonable prediction of screech-tone frequencies. A similar resonant mechanism was proposed for subsonic compressible jets \citep{towne2017acoustic}, cavity flows \citep{rossiter1964wind, rowley2002self} and impinging jets \citep{tam1990theoretical, bogey2017feedback}. The reflection of waves is implicitly considered in all these mechanisms, but it is rarely studied in detail. In more complete screech-frequency prediction models \citep{mancinelli2019reflection, mancinelli2021complexvalued}, where the spatial growth rates of the upstream- and downstream-travelling waves are included, knowledge of the reflection coefficients in the jet exit plane and at the location of STI is required. 

In this paper, we investigate the interaction between a downstream-travelling K-H wave and a normal shock and compute the amplitude and phase of the reflected upstream-travelling guided wave active in the screech loop using a mode-matching approach. We consider vortex-sheet (V-S) and finite-thickness (F-T) flow models, which elucidates the role of shear in the reflection and transmission processes. The efficiency of the mode-matching technique in the presence of a discontinuity, as is the shock in the flow we consider herein, has been already shown by \cite{gabard2008computational} for the estimation of the sound attenuation in a lined duct. More recently, a mode-matching approach has been used by \cite{dai2020flow} and \cite{dai2021total} to calculate the reflection and transmission coefficients in a duct flow in the presence of a cavity. Consistent with what was done in these works, we use linear theory to describe the flow dynamics upstream and downstream of the shock and then match the solutions across the discontinuity. The paper is organised as follows. The general modelling framework is presented in \S\ref{sec:model}. Results are presented and discussed in \S\ref{sec:results}. The paper closes with concluding remarks in \S\ref{sec:conclusions}.

\section{Modelling framework}
\label{sec:model}
\begin{figure}
\centering
\subfigure[]{\includegraphics[scale=0.7]{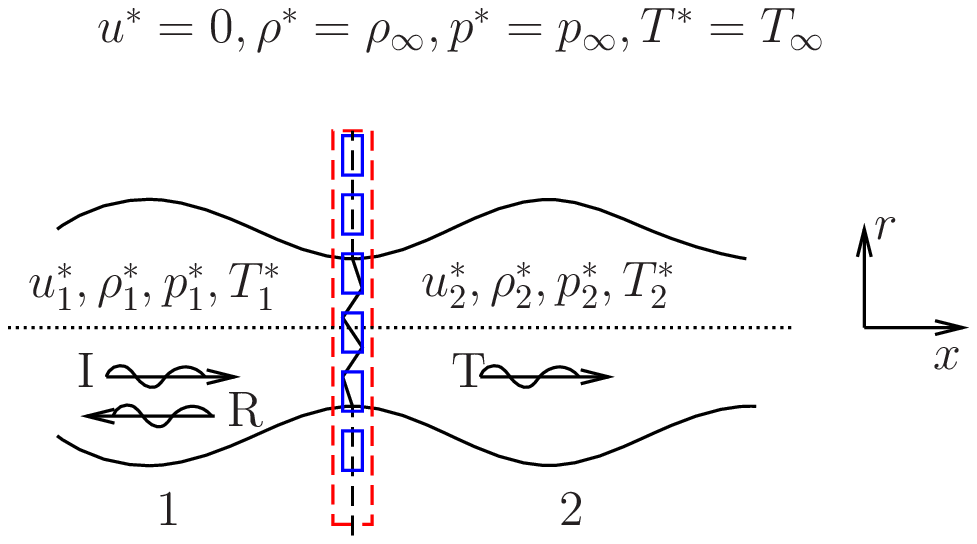}}
\subfigure[]{\includegraphics[scale=0.33]{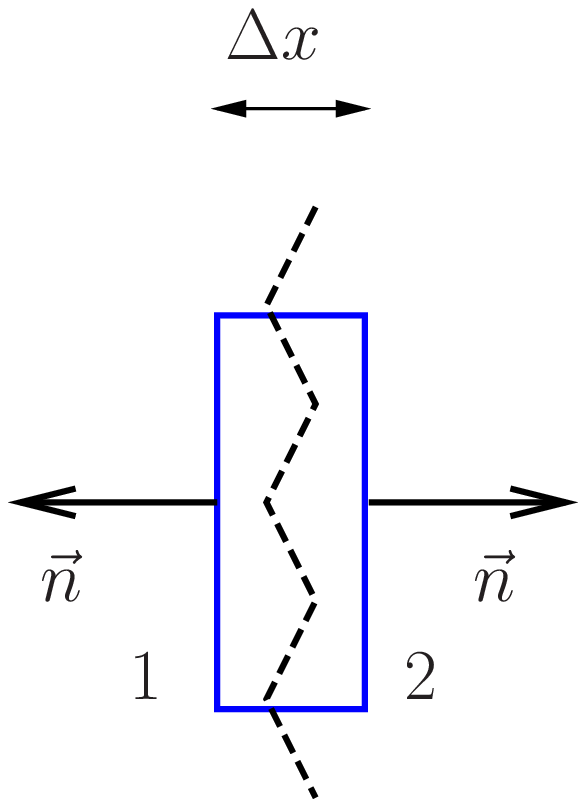}}
\caption{Schematic representation of the jet model: (a) sketch of the shock-containing jet, (b) control volume representation with identification of the normal to the inlet and outlet surfaces.}
\label{fig:jet_model}
\end{figure}

We here present the shock and jet-dynamics modelling and the procedure adopted to calculate the reflection and transmission coefficients. We consider an axisymmetric, shock-containing supersonic jet. With the aim of keeping the model as simple as possible, the shock is assumed to be normal, thus allowing the use of the locally parallel-flow assumption both upstream and downstream of the shock. A sketch of the shock-containing jet and the cylindrical reference system used in this paper are depicted in figure \ref{fig:jet_model}. We assume a K-H wave with unitary amplitude, $I=1$, incident to a shock. The interaction of the incoming wave with the shock generates a collection of reflected and transmitted modes upstream and downstream of the shock, respectively. The sections upstream and downstream of the shock are hereinafter denoted 1 and 2 and the reflection and transmission coefficients of each wave moving away from the shock are indicated with $R_{n_R}$ and $T_{n_T}$, respectively. The state vector is $\mathbf{q}^*=\left\lbrace\rho^*,\,u_x^*,\,u_r^*,\,u_\theta^*,\,T^*,\,p^*\right\rbrace$. The flow variables are normalised by the nozzle diameter $D$, the ambient density and speed of sound $\rho_\infty$ and $c_\infty$, respectively, thus leading to a non-dimensional state vector $\mathbf{q}$.

\subsection{Shock model}
\label{subsec:shock_model}
The flow regions upstream and downstream of the shock are well described by a locally-parallel model. In order to connect these two regions, we impose the conservation laws of mass, momentum and energy through the shock. This is done by dividing the shock into infinitesimal control volumes $\mathrm{d}V$ of length $\Delta x\to 0$ such that the flux terms through the top and bottom surfaces are zero (see figure \ref{fig:jet_model}) and enforcing mass, momentum, and energy conservation for the control volume, leading to the system of equations

\begin{subnumcases}{\label{eq:conservation}}
\int\limits_{S_1}\rho^*\vec{u}^*\cdot\vec{n}\,\mathrm{d}S + \int\limits_{S_2}\rho^*\vec{u}^*\cdot\vec{n}\,\mathrm{d}S=0\mathrm{,}\\
\int\limits_{S_1}\rho^*\vec{u}^*\left(\vec{u}^*\cdot\vec{n}\right)\,\mathrm{d}S + \int\limits_{S_1} p^*\vec{n}\,\mathrm{d}S + \int\limits_{S_2}\rho^*\vec{u}^*\left(\vec{u}^*\cdot\vec{n}\right)\,\mathrm{d}S - \int\limits_{S_2} p^*\vec{n}\,\mathrm{d}S=0\mathrm{,}\\
\int\limits_{S_1}\rho^* e^*\vec{u}^*\cdot\vec{n}\,\mathrm{d}S + \int\limits_{S_2}\rho^* e^*\vec{u}^*\cdot\vec{n}\,\mathrm{d}S=0\mathrm{,}
\end{subnumcases}

\noindent where $e^*=h^*+0.5\left(u_x^{*2}+u_r^{*2}+u_\theta^{*2}\right)$ is the specific energy with the enthalpy expressed as $h^*=c_pT^*$. Normalising the flow variables, performing the Reynolds decomposition,

\begin{equation}
\mathbf{q}\left(x,r,\theta,t\right)=\overline{\mathbf{q}}\left(r\right)+\mathbf{q}'\left(x,r,\theta,t\right)\mathrm{,}
\label{eq:Re_decomposition}
\end{equation}

\noindent and substituting into \eqref{eq:conservation}, removing the mean and linearising, the linearised jump equations for the shock become

\begin{subnumcases}{\label{eq:linear_shock}}
\overline{u}_{1x}\rho_1+\overline{\rho}_1u_{1x}=\overline{u}_{2x}\rho_2+\overline{\rho}_2u_{2x}\mathrm{,}\\
p_1+2\overline{\rho}_1\overline{u}_{1x}u_{1x}+\overline{u}_{1x}^2\rho_1=p_2+2\overline{\rho}_2\overline{u}_{2x}u_{2x}+\overline{u}_{2x}^2\rho_2\mathrm{,}\\
u_{1r}=u_{2r}\mathrm{,}\\
u_{1\theta}=u_{2\theta}\mathrm{,}\\
T_1+\overline{u}_{1x}u_{1x}=T_2+\overline{u}_{2x}u_{2x}\mathrm{,}
\end{subnumcases}

\noindent where we removed the primes from the fluctuating variables for notational simplicity. The perturbations upstream and downstream of the shock are modelled using the normal mode ansatz

\begin{equation}
\mathbf{q}\left(x,r,\theta,t\right)=\hat{\mathbf{q}}\left(r\right)e^{i\left(kx+m\theta -\omega t\right)}\mathrm{,}
\label{eq:normal_ansatz}
\end{equation}

\noindent where $k$ is the wavenumber along the axial direction, $m$ is the azimuthal order and $\omega=2\pi St M_a$ is a non-dimensional frequency, with $St=fD/U_j$ the nozzle-diameter-based Strouhal number and $M_a=U_j/c_\infty$ the acoustic Mach number. Considering that the only incident wave is the K-H wave, the system of equations \eqref{eq:linear_shock} can be written in a compact form as follows,

\begin{equation}
\mathbf{A}_1\left(I\hat{\mathbf{q}}_{1I}+\sum\limits_{n_R=1}^\infty R_{n_R}\hat{\mathbf{q}}_{1R,n_R}\right)=\mathbf{A}_2\sum\limits_{n_T=1}^\infty T_{n_T}\hat{\mathbf{q}}_{2T,n_T}
\mathrm{,}
\label{eq:balance}
\end{equation}

\noindent where $\hat{\mathbf{q}}_{1I}$, $\hat{\mathbf{q}}_{1R,n_R}$ and $\hat{\mathbf{q}}_{2T,n_T}$ are the the incident, reflected and transmitted waves moving upstream and downstream of the shock, respectively, and

\begin{equation}
\mathbf{A}_1=
\begin{bmatrix}
\overline{u}_{1x} & \overline{\rho}_1 & 0 & 0 & 0 & 0\\
\overline{u}_{1x}^2 & 2\overline{\rho}_1\overline{u}_{1x} & 0 & 0 & 0 & 1\\
0 & 0 & 1 & 0 & 0 & 0\\
0 & 0 & 0 & 1 & 0 & 0\\
0 & \overline{u}_{1x} & 0 & 0 & 1 & 0
\end{bmatrix}
\mathrm{,}\qquad
\mathbf{A}_2=
\begin{bmatrix}
\overline{u}_{2x} & \overline{\rho}_2 & 0 & 0 & 0 & 0\\
\overline{u}_{2x}^2 & 2\overline{\rho}_2\overline{u}_{2x} & 0 & 0 & 0 & 1\\
0 & 0 & 1 & 0 & 0 & 0\\
0 & 0 & 0 & 1 & 0 & 0\\
0 & \overline{u}_{2x} & 0 & 0 & 1 & 0
\end{bmatrix}
\end{equation}

\noindent are matrices containing information about the mean flow upstream and downstream of the shock. The vector of eigenfunctions $\hat{\mathbf{q}}_{1I}$, $\hat{\mathbf{q}}_{1R,n_R}$ and $\hat{\mathbf{q}}_{2T,n_T}$, are computed using either a vortex sheet or finite-thickness model (see \S\ref{subsec:jet_modelling}). The procedure used to ascertain whether a wave is reflected or transmitted is described in \S\ref{subsubsec:travel_direction}.

\subsection{Reflection- and transmission-coefficients calculation}
\label{subsec:R_T_computation}
Equation \eqref{eq:balance} is exact if a complete basis of jet modes is considered. In order to estimate the reflection- and transmission-coefficient values, we truncate the sum to a finite number of modes $N_R$ and $N_T$ and we introduce an error density $\epsilon\left(r\right)$ for each conservation equation. Equation \eqref{eq:balance} can then be written as

\begin{equation}
\mathbf{A}_1\left(I\hat{\mathbf{q}}_{1I}+\sum\limits_{n_R=1}^{N_R} R_{n_R}\hat{\mathbf{q}}_{1R,n_R}\right)-\mathbf{A}_2\sum\limits_{n_T=1}^{N_T} T_{n_T}\hat{\mathbf{q}}_{2T,n_T}=\bm{\epsilon}
\mathrm{,}
\label{eq:balance_truncate}
\end{equation}

\noindent where $\bm{\epsilon}\left(r\right)$ is the error density vector and $\bm{\epsilon}\to 0$ if the number of modes $N_R$ and $N_T\to \infty$. The reflection and transmission coefficients $R_{n_R}$ and $T_{n_T}$ associated with each mode are estimated by a least-mean-square minimization of the error densities, as formalised in the following,

\begin{equation}
\left[R_{opt,n_R},T_{opt,n_T}\right]=\underset{R_{n_R},T_{n_T} \in\,\mathcal{C}}{\text{min}}\underbrace{\int\limits_0^\infty\vert\bm{\epsilon}^2\left(r\right)\vert\,\mathrm{d}r}_{F}\mathrm{.}
\label{eq:minimization}
\end{equation}

The objective function $F$ corresponds to the sum of the absolute value of the squared error densities associated with each conservation equation integrated along the radial direction. To solve the minimization problem in \eqref{eq:minimization}, we write \eqref{eq:balance_truncate} in a matrix form,

\begin{equation}
\begin{bmatrix}
\mathbf{A}_1\hat{\mathbf{Q}}_{1R}\quad -\mathbf{A}_2\hat{\mathbf{Q}}_{2T}
\end{bmatrix}
\begin{Bmatrix}
\mathbf{R}\\ \mathbf{T}
\end{Bmatrix}
=-I\mathbf{A}_1\hat{\mathbf{q}}_{1I}+\bm{\epsilon}\mathrm{.}
\label{eq:error_matrix}
\end{equation}

\noindent where $\hat{\mathbf{Q}}_{1R}=\left[\hat{\mathbf{q}}_{1R,1},\,...,\,\hat{\mathbf{q}}_{1R,N_R}\right]$ and $\hat{\mathbf{Q}}_{2T}=\left[\hat{\mathbf{q}}_{2T,1},\,...,\,\hat{\mathbf{q}}_{2T,N_T}\right]$ are the matrices of the eigenfunctions and $\mathbf{R}$ and $\mathbf{T}$ are the vectors of the reflection and transmission coefficients, respectively. We then define,

\begin{equation}
\begin{Bmatrix}
\mathbf{R}\\
\mathbf{T}
\end{Bmatrix}
= \mathbf{X}\qquad
\begin{bmatrix}
\mathbf{A}_1\hat{\mathbf{Q}}_{1R} \quad -\mathbf{A}_2\hat{\mathbf{Q}}_{2T}
\end{bmatrix}
=\mathbf{B}
\qquad \mathbf{y}=\mathbf{A}_1\hat{\mathbf{q}}_{1I}I\mathrm{,}
\label{eq:defi}
\end{equation}

\noindent so that \eqref{eq:error_matrix} can be written in the compact form $\bm{\epsilon}=\mathbf{B}\mathbf{X}+\mathbf{y}$. The solution of \eqref{eq:minimization} is obtained by finding the stationary point of $F$ by setting its gradient to zero, 

\begin{equation}
\frac{dF}{d\mathbf{X}}=\frac{d\left(\bm{\epsilon^T}\mathbf{W}\bm{\epsilon}\right)}{d\mathbf{X}} = \mathbf{B}^T\mathbf{W}\mathbf{B}\mathbf{X}+\mathbf{B}^T\mathbf{W}\mathbf{y}=0\mathrm{,}
\end{equation}

\noindent which leads to,

\begin{equation}
\mathbf{X}=-\left(\mathbf{B}^T\mathbf{W}\mathbf{B}\right)^{-1}\mathbf{B}^T\mathbf{W}\mathbf{y}\mathrm{,}
\label{eq:pinv}
\end{equation}

\noindent where $\mathbf{B}\vert_W^+=\left(\mathbf{B}^T\mathbf{W}\mathbf{B}\right)^{-1}\mathbf{B}^T\mathbf{W}$ is the weighted pseudo-inverse matrix and $\mathbf{W}$ is a diagonal matrix of elements $dr$.

\subsection{Jet models}
\label{subsec:jet_modelling}
We here present a local description of the jet dynamics using the parallel-flow linear stability theory. This theory is applied to two different models: a finite-thickness flow model and a simplified cylindrical vortex sheet. Both are governed by the Linearised Euler Equations (LEE) (see \S\ref{sec:Euler_app}).

\subsubsection{Finite-thickness model}
\label{subsubsec:shear_layer}
Writing the LEE with respect to the pressure, the compressible Rayleigh equation

{\small
\begin{equation}
\frac{\partial^2\hat{p}}{\partial r^2}+\left(\frac{1}{r}-\frac{2k}{\overline{u}_xk-\omega}\frac{\partial\overline{u}_x}{\partial r}-\frac{\gamma -1}{\gamma\overline{\rho}}\frac{\partial\overline{\rho}}{\partial r}+\frac{1}{\gamma\overline{T}}\frac{\partial\overline{T}}{\partial r}\right)\frac{\partial\hat{p}}{\partial r}-\left(k^2+\frac{m^2}{r^2}-\frac{\left(\overline{u}_xk-\omega\right)^2}{\left(\gamma -1\right)\overline{T}}\right)\hat{p}=0\mathrm{,}
\label{eq:Rayleigh}
\end{equation}}

\noindent is obtained, where $\gamma$ is the specific heat ratio for a perfect gas. The solution of the linear stability problem is obtained specifying a real or complex frequency $\omega$ and solving the resulting augmented eigenvalue problem $k=k\left(\omega\right)$, with $\hat{p}\left(r\right)$ the associated pressure eigenfunction. The eigenvalue problem is solved numerically by discretising \eqref{eq:Rayleigh} in the radial direction using the Chebyshev polynomials. A mapping function proposed by \cite{lesshafft2007linear} was used to non-uniformly distribute the 500 grid points to efficiently resolve the shear layer of the jet. The eigenfunctions $\hat{u}_i$, $\hat{\rho}$ and $\hat{T}$ are calculated from the knowledge of $\hat{p}$ (see \S\ref{sec:Euler_app}). The eigenfunctions are normalised such that $\angle\hat{p}\left(r\right)=0$ for $r=0$ and to have unitary energy norm, which, following \cite{chu1965energy} and \cite{hanifi1996transient}, is defined as

\begin{equation}
E=\frac{1}{2}\int\limits_0^{2\pi}\int\limits_0^\infty \left(\overline{\rho}\left(\vert \hat{u_x}\vert^2+\vert \hat{u_r}\vert^2+\vert \hat{u_\theta}\vert^2\right) + \frac{\gamma -1}{\gamma}\frac{\overline{T}}{\overline{\rho}}\vert\hat{\rho}\vert^2 + \frac{\overline{\rho}}{\gamma\overline{T}}\vert \hat{T}\vert^2\right)r\,\mathrm{d}r\,\mathrm{d}\theta\mathrm{.}
\label{eq:energy_norm}
\end{equation}

The derivation of the energy norm is provided in \S\ref{sec:energy_norm_app}.

\subsubsection{Vortex-sheet model}
\label{subsubsec:VS}
The vortex-sheet model is an inviscid idealisation of the jet where the infinitely thin vortex sheet separates the interior flow and the outer quiescent fluid, resulting in a jet with a mean top-hat profile. The vortex sheet was used by \cite{lessen1965inviscid} and \cite{michalke1970note} to study the stability properties of a compressible jet. We recently showed that the standard V-S model for free jets, which was used in the previous studies, does not support free-stream acoustic modes \citep{mancinelli2021completing}. We, thus, considered the dispersion relation of a confined jet with the radial distance of the boundary sufficiently distant from the jet in order to recover the same dynamical properties of a free jet. The analysis of this surrogate problem allowed us to include the free-stream acoustic modes in the description of the jet dynamics (for more details the reader can refer to \cite{mancinelli2021completing}). We herein use this confined version of the vortex sheet, whose dispersion relation $D\left(k,\omega;M_a,T,m,r_{MAX}\right)=0$ is,

\begin{equation}
\begin{split}
&\frac{1}{\left(1-\frac{kM_a}{\omega}\right)^2} + \frac{1}{T}\frac{I_m\left(\frac{\gamma_i}{2}\right)}{K_m\left(\frac{\gamma_o}{2}\right)-zI_m\left(\frac{\gamma_o}{2}\right)}\\
&\frac{\frac{\gamma_o}{2}K_{m-1}\left(\frac{\gamma_o}{2}\right) + mK_m\left(\frac{\gamma_o}{2}\right)+z\left(\frac{\gamma_o}{2}I_{m-1}\left(\frac{\gamma_o}{2}\right)-mI_m\left(\frac{\gamma_o}{2}\right)\right)}{\frac{\gamma_i}{2}I_{m-1}\left(\frac{\gamma_i}{2}\right) - mI_m\left(\frac{\gamma_i}{2}\right)}=0\mathrm{,}
\end{split}
\label{eq:dispersion}
\end{equation}

\noindent with

\begin{subequations}
\begin{align}
&\gamma_i = \sqrt{k^2-\frac{1}{T}\left(\omega-M_ak\right)^2}\\
&\gamma_o = \sqrt{k^2-\omega^2}\mathrm{,}
\end{align}
\label{eq:gamma}
\end{subequations}

\noindent where $I$ and $K$ are modified Bessel functions of the first and second kind, respectively, $T=T_j/T_\infty$ is the jet-to-ambient temperature ratio such that the relation between the jet and the acoustic Mach numbers is $M_a=M_j\sqrt{T}$, and $z=K_m\left(\gamma_or_{MAX}\right)/I_m\left(\gamma_or_{MAX}\right)$. The branch cut in the square root of \eqref{eq:gamma} is chosen such that $-\pi/2\leq \rm{arg}\left(\gamma_{i,o}\right)<\pi/2$. We remind the reader that the dispersion relation in \eqref{eq:dispersion} for a confined jet differs from the standard one for a free jet in the additional terms containing $z\left(r_{MAX}\right)$. Following \cite{mancinelli2021completing}, we herein use $r_{MAX}=100$ in order to avoid any effect of the boundary on the eigenmodes. Frequency/wavenumber pairs $\left(\omega, k\right)$ that satisfy \eqref{eq:dispersion} define eigenmodes of the vortex sheet for given values of $m$, $M_a$, $T$ and $r_{MAX}$. To find these pairs, similarly to the finite-thickness model, we specify a frequency $\omega$ (real or complex) and compute the associated eigenvalues $k$. Eigenvalues are computed using a root finder based on the Levenberg-Marquardt method \citep{levenberg1944method, marquardt1963algorithm}.

The solution for the pressure in the inner and outer flows is provided by,

\begin{subnumcases}{\label{eq:VS_p}}
\hat{p}_i\left(r\right)=B_iI_m\left(\gamma_ir\right) \qquad r\leq 0.5\\
\hat{p}_o\left(r\right)=C_o\left(-zI_m\left(\gamma_or\right) + K_m\left(\gamma_or\right)\right) \qquad r> 0.5\mathrm{,}
\end{subnumcases}

\noindent where $B_i$ and $C_o$ are constants fixed in order to ensure pressure continuity at the vortex sheet location $r=0.5$.

The eigenfunctions of the other flow variables are calculated from the knowledge of $\hat{p}_{i,o}\left(r\right)$ by exploiting the Fourier-transformed LEE \eqref{eq:F_LEE}. The same eigenfunction normalisation procedure described for the finite-thickness model is used for the vortex-sheet model as well.

\subsubsection{Identification of reflected and transmitted waves}
\label{subsubsec:travel_direction}
There are two types of waves which appear due to the scattering of the K-H wave at the shock: upstream-travelling reflected waves in region 1, and downstream-travelling transmitted waves in region 2. Following \cite{towne2017acoustic}, we use the terms downstream- and upstream-travelling to designate the direction of the energy transfer. This property can be characterised using the Briggs-Bers criterion by looking at the asymptotic behaviour of $k\left(\omega\right)$ at large $\omega_i$ \citep{briggs1964electron, bers1983space}. The wave is downstream-travelling if

\begin{subequations}
\begin{equation}
\lim_{\omega_i\to+\infty}k_i=+\infty
\end{equation}

\noindent and upstream-travelling if 

\begin{equation}
\lim_{\omega_i\to+\infty}k_i=-\infty\mathrm{,}
\end{equation}
\label{eq:briggs}
\end{subequations}

\noindent where the subscript $i$ stands for the imaginary part of the variable. The downstream- and upstream-travelling waves are denoted hereinafter with the superscript $+$ and $-$, respectively.

\subsubsection{Mean flow}
\label{subsubsec:mean_flow_LEE}
The conditions upstream of the shock in the case of the vortex sheet are provided by,

\begin{subequations}
\begin{minipage}{0.49\textwidth}
\begin{equation}
r\leq 0.5\;
\begin{cases}
\overline{u}_{1x}=M_{a_1}\\
\overline{p}_1=\frac{\rho T}{\gamma}\\
\overline{\rho}_1=\rho\\
\overline{T}_1=\frac{T}{\gamma-1}
\end{cases}
\label{eq:mean_VS_1_in}
\end{equation}
\end{minipage}
\begin{minipage}{0.48\textwidth}
\begin{equation}
\qquad r>0.5\;
\begin{cases}
\overline{u}_{1x}=0\\
\overline{p}_1=\frac{1}{\gamma}\\
\overline{\rho}_1=1\\
\overline{T}_1=\frac{1}{\gamma-1}\mathrm{,}
\end{cases}
\label{eq:mean_VS_1_out}
\end{equation}
\end{minipage}
\end{subequations}

\noindent where $\rho=\rho_j/\rho_\infty$ is the jet-to-ambient density ratio.

In the case of finite thickness, we use the hyperbolic tangent function reported in \cite{lesshafft2007linear} for the velocity profile upstream of the shock,

\begin{equation}
\overline{u}_{1x}=\frac{1}{2}M_{a_1}\left(1+\tanh\left(\frac{R}{4\theta_R}\left(\frac{R}{r}-\frac{r}{R}\right)\right)\right)\mathrm{,}
\label{eq:hyperbolic}
\end{equation}

\noindent where $\theta_R$ is the shear-layer momentum thickness and $R=0.5$ is the nozzle radius. Consistent with Particle Image Velocimetry results presented in \cite{mancinelli2021complexvalued} for an under-expanded supersonic jet with jet Mach number $M_{j_1}=1.1$ and temperature ratio $T\approx 0.81$, we choose a shear-layer thickness $R/\theta_R=10$.

The mean flow downstream of the shock can then be determined from the upstream conditions using the jump equations of normal shocks,

\begin{subequations}
\begin{empheq}[left=\empheqlbrace]{align}
& M_{j_2}=\sqrt{\frac{M_{j_1}^2\left(\gamma-1\right)+2}{2\gamma M_{j_1}^2-\left(\gamma-1\right)}}\mathrm{,}\\
& \frac{\overline{p}_2}{\overline{p}_1}=\frac{2\gamma M_{j_1}^2-\left(\gamma-1\right)}{\gamma+1}\mathrm{,}\\
& \frac{\overline{\rho}_2}{\overline{\rho}_1}=\dfrac{\left(\gamma+1\right)M_{j_1}^2}{\left(\gamma-1\right)M_{j_1}^2+2}\mathrm{,}\\
& \frac{\overline{T}_2}{\overline{T}_1}=\frac{\left(1+\frac{\gamma-1}{2}M_{j_1}^2\right)\left(\frac{2\gamma}{\gamma-1}M_{j_1}^2-1\right)}{M_{j_1}^2\left(\frac{2\gamma}{\gamma-1}+\frac{\gamma-1}{2}\right)}\mathrm{.}
\end{empheq}
\label{eq:jump}
\end{subequations} 

\begin{figure}
\centering
\subfigure[]{\includegraphics[scale=0.27]{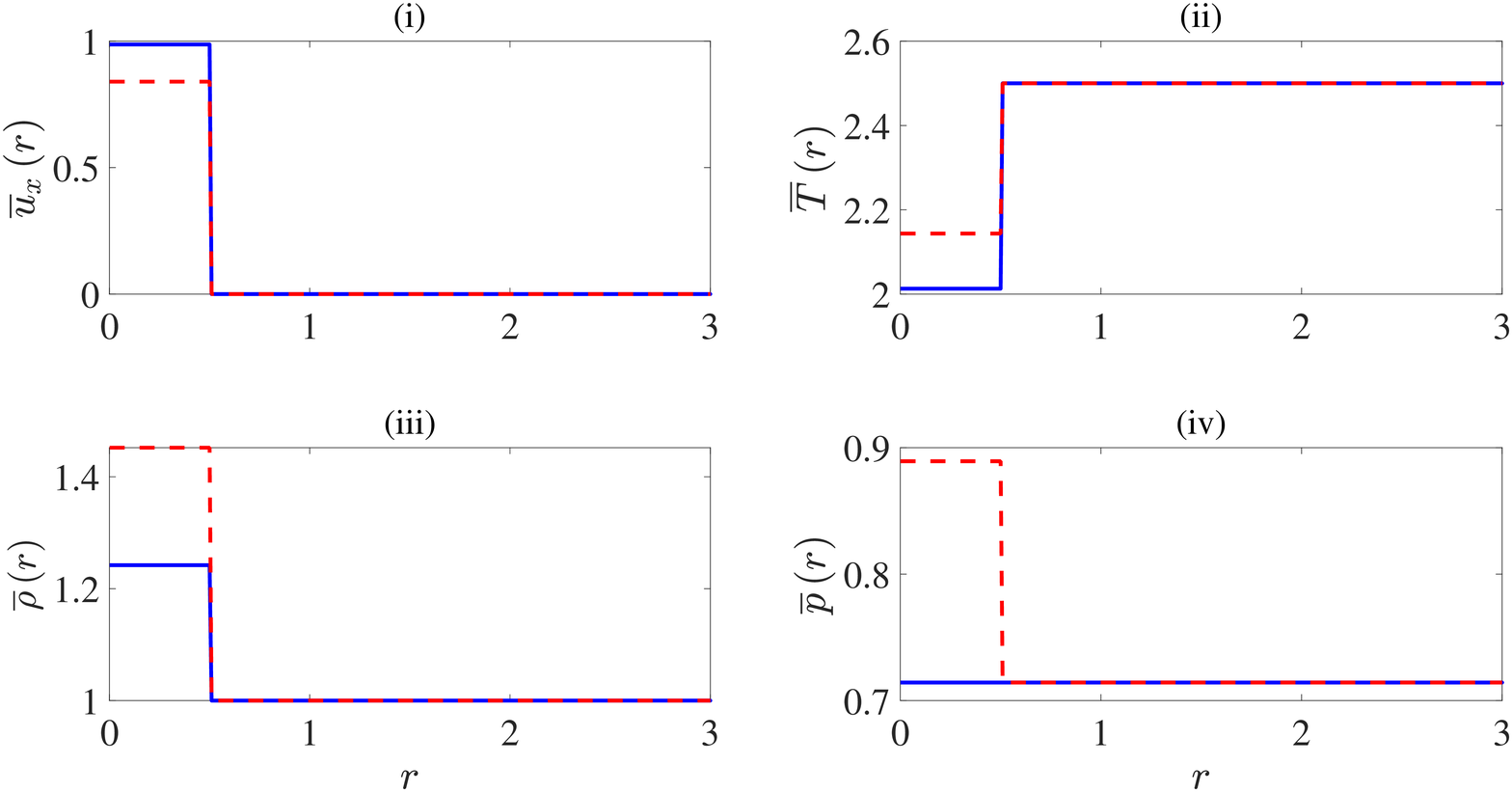}}
\subfigure[]{\includegraphics[scale=0.27]{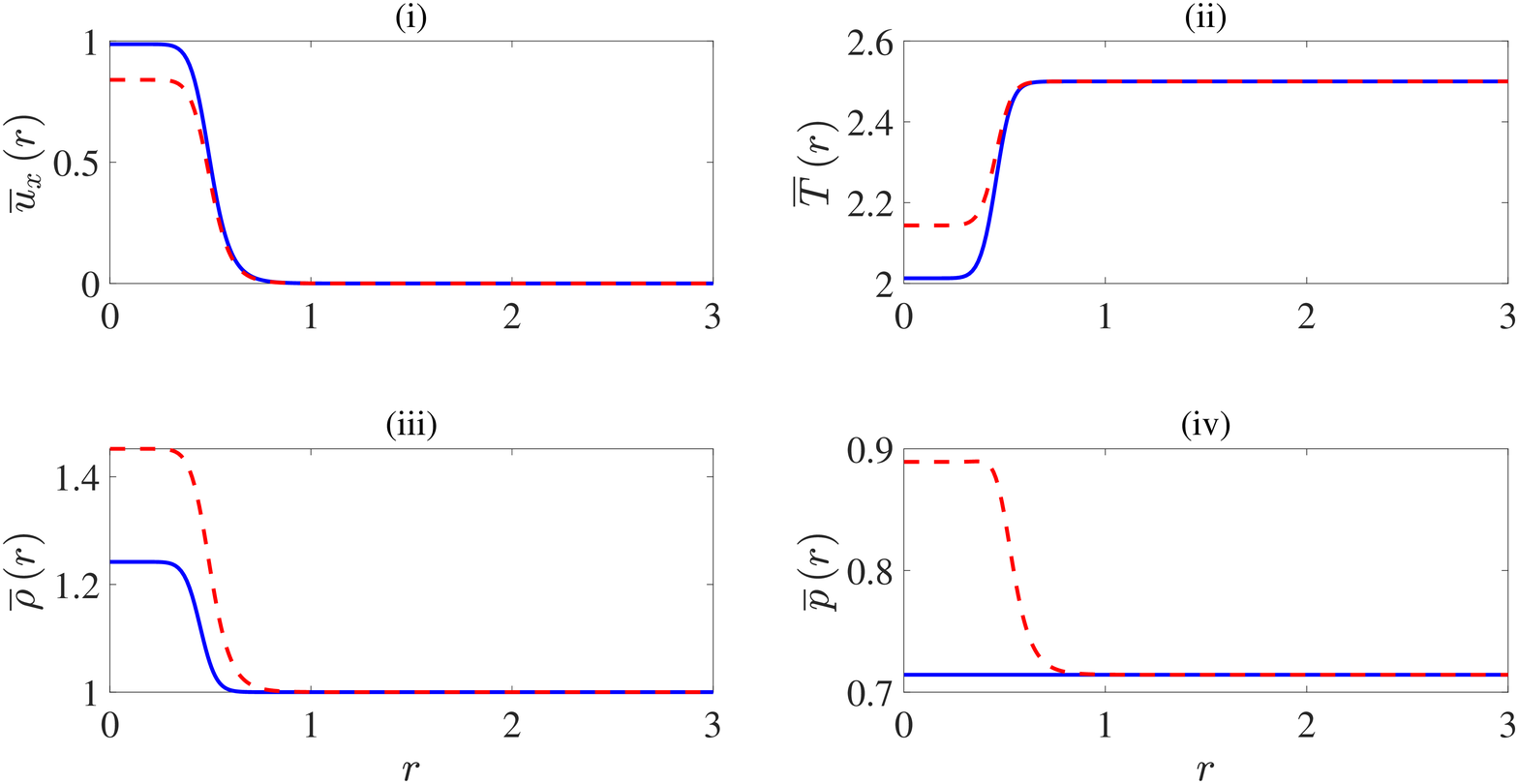}}
\caption{Mean-flow upstream and downstream of the shock for the vortex-sheet and finite-thickness models for $M_{j_1}=1.1$: solid blue lines refer to upstream conditions, dashed red lines to downstream ones. (a) V-S model, (b) finite-thickness model: (i) axial velocity, (ii) temperature, (iii) density, (iv) pressure.}
\label{fig:mean}
\end{figure}

The mean flow profiles upstream and downstream of the shock in the case of vortex-sheet and finite-thickness models for the flow conditions listed above are represented in figure \ref{fig:mean}. The presence of the shock wave generates a mean pressure gradient along the radial direction downstream of the shock. This $\partial\overline{p}/\partial r$ induces the appearance of a mean radial velocity $\overline{u}_r$, thus making the flow slowly diverging. The evaluation of this induced radial velocity in the case of a finite-thickness model is reported in appendix \ref{sec:radial_mean}, where we show that the induced mean radial velocity is small in comparison with the axial velocity component. We also point out that the transmission and reflection mechanisms occur locally and hence they are not affected by the flow evolution far away from the shock.

\section{Results}
\label{sec:results}
We here present the results of the reflection-coefficient calculation obtained by modelling the jet dynamics with both the vortex-sheet and finite-thickness models. We first consider a Strouhal number $St=0.68$, a jet Mach number $M_{j_1}=1.1$ and a temperature ratio $T\approx 0.81$ (corresponding to an acoustic Mach number $M_{a_1}\approx 0.987$), which results in a jet Mach number $M_{j_2}=0.91$ and an acoustic Mach number $M_{a_2}\approx 0.84$ downstream of the shock. These upstream jet conditions and Strouhal number are selected to match conditions for which screech has been experimentally observed by \cite{mancinelli2019screech} and \cite{mancinelli2021complexvalued}. Due to the axisymmetric nature of the resonance mode, we here study the azimuthal mode $m=0$. Furthermore, we focus on the reflection coefficient of the upstream-travelling guided mode of the second radial order given that this mode has been proved to be the closure mechanism for axisymmetric screech modes (see \cite{edgington2018upstream}, \cite{mancinelli2019screech} and \cite{mancinelli2021complexvalued}). Finally, for the finite-thickness model, we explore the variation of the reflection coefficient as a function of both $St$ and $M_j$.

\subsection{Vortex-sheet model}
\label{subsec:VS_results}
\begin{figure}
\centering
\includegraphics[scale=0.27]{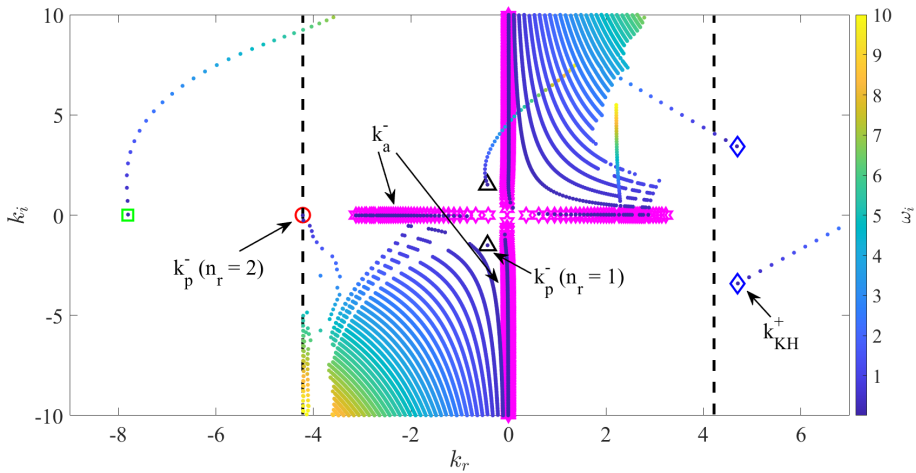}
\caption{Eigenspectrum upstream of the shock for azimuthal mode $m=0$, $T\approx 0.81$ and $M_{j_1}=1.1$. Empty markers represent eigenvalues from the vortex-sheet model for $\omega\in\mathcal{R}$, $\bullet$ for $\omega\in\mathcal{C}$, with $\omega_i\to+\infty$ from blue to yellow. Blue $\diamond$ represent $k_{KH}^+$ and $k_{KH}^{*+}$ waves, red $\circ$ represents propagative $k_p^-$ mode with $n_r=2$, black $\bigtriangleup$ represent evanescent $k_p^\pm$ modes with $n_r=1$, green $\Box$ represent $k_p^+$ modes with $n_r\geq 2$. Magenta $\ast$ represent $k_a^\pm$ waves. Dashed lines refer to the sonic speed $\pm c_\infty$.}
\label{fig:eigenspectrum_VS_up}
\end{figure}

Figure \ref{fig:eigenspectrum_VS_up} shows the eigenspectrum in the $k_r$-$k_i$ plane upstream of the shock for both real and complex $\omega$. Distinct families of modes can be identified. The V-S model supports one convectively unstable mode, the K-H mode, which is denoted hereinafter $k_{KH}$. The unstable $k_{KH}$ wave has a complex-conjugate and both eigenvalues have positive phase and group velocities according to the criteria \eqref{eq:briggs}. The V-S model also supports guided modes, i.e., modes that use the jet as a wave guide. These modes, hereinafter denoted $k_p$, belong to a hierarchical family of waves identified by their azimuthal and radial orders $m$ and $n_r$, respectively. According to \cite{towne2017acoustic}, these modes are guided or completely trapped inside the jet depending on the $St$ and the radial order considered. Specifically, we observe evanescent $k_p$ waves for $n_r=1$ with supersonic negative phase speed. The wave associated with $k_i>0$ is a downstream-travelling wave, whereas the wave with $k_i<0$ is upstream-travelling. The $k_p$ mode for $n_r = 2$ is propagative and upstream-travelling and has a slightly subsonic negative phase speed. All the $k_p^\pm$ modes with $n_r\leq 2$ have support both inside and outside of the jet for the $St$ analysed. We may then identify all the $k_p^+$ modes for $n_r> 2$. These modes represent acoustic waves trapped inside the jet due to total reflection at the vortex sheet, which effectively behaves as a soft-walled duct \citep{towne2017acoustic,martini_cavalieri_jordan_2019}. Finally, we find propagative and evanescent acoustic modes, which are hereinafter denoted $k_a$. Among them, modes lying on the real and imaginary axes with $k_r<0$ and $k_i<0$, respectively, are upstream-travelling modes. To summarise, the vortex-sheet model supports two families of reflected waves upstream of the shock: (i) guided jet modes and (ii) propagative and evanescent acoustic modes. Among the guided modes, we may distinguish the evanescent $k_p^-$ mode with $n_r=1$ and the propagative $k_p^-$ mode with $n_r=2$. In the remainder of this paper, we focus our attention on the reflection coefficient between the KH wave and the propagative $k_p^-$ mode with $n_r = 2$, since this interaction is responsible for the screech resonance at this frequency and Mach number. 

\begin{figure}
\centering
\subfigure[]{\includegraphics[scale=0.27]{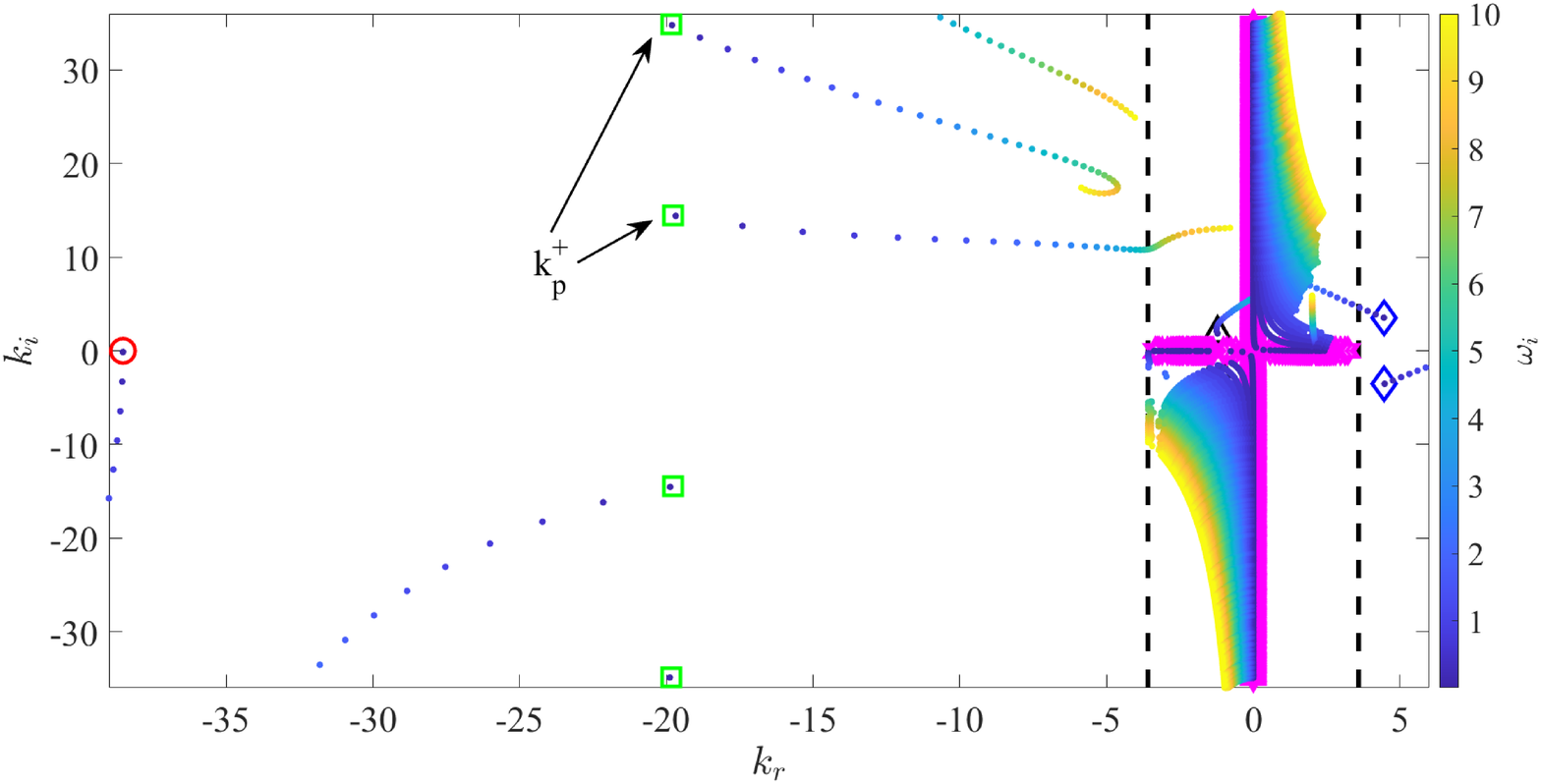}}
\subfigure[]{\includegraphics[scale=0.27]{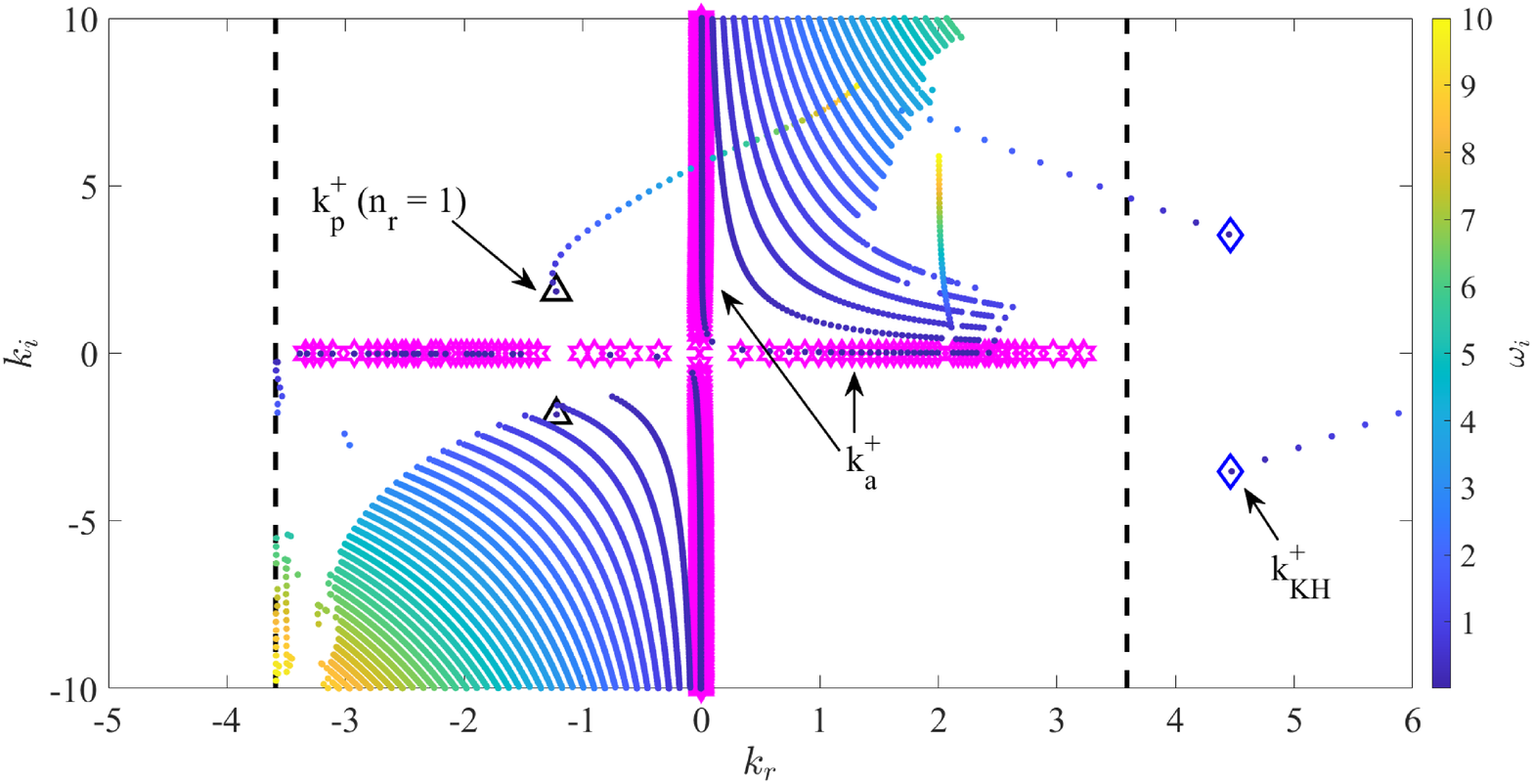}}
\caption{Eigenspectrum downstream of the shock for azimuthal mode $m=0$, $T\approx 0.85$ and $M_{j_2}=0.91$ which corresponds to $M_{a_2}=0.84$: (a) eigenspectrum, (b) zoom around the origin. Empty markers represent eigenvalues from vortex-sheet model for $\omega\in\mathcal{R}$, $\bullet$ for $\omega\in\mathcal{C}$,with $\omega_i\to+\infty$ from blue to yellow. Blue $\diamond$ represent $k_{KH}^+$ and $k_{KH}^{*+}$ waves, red $\circ$ represents propagative $k_d^-$ mode with $n_r=1$, black $\bigtriangleup$ represent evanescent $k_p^\pm$ modes with $n_r=1$, green $\Box$ represent $k_p^\pm$ modes with $n_r\geq 2$. Magenta $\ast$ represent $k_a^\pm$ waves. Dashed lines refer to the sonic speed $\pm c_\infty$.}
\label{fig:eigenspectrum_VS_down}
\end{figure}

We now consider the eigenspectrum downstream of the shock in figure \ref{fig:eigenspectrum_VS_down} to identify the transmitted modes. According to \cite{towne2017acoustic}, in the subsonic regime the guided modes are characterised by two upstream-travelling branches delimited by two saddle points in the $k_r$-$St$ plane: one branch characterised by a larger negative phase speed close to the ambient speed of sound, herein denoted $k_p^-$, and one with a lower phase-speed absolute value, herein denoted $k_d^-$. Similar to the eigenspectrum upstream of the shock, we may identify the $k_{KH}^+$ wave and its complex-conjugate $k_{KH}^{*+}$, the downstream- and upstream-travelling $k_p$ waves with $n_r=1$, which are evanescent and have supersonic negative phase speed at this frequency, and the $k_a^\pm$ modes. We then locate the upstream-travelling propagative $k_d^-$ wave for $n_r=1$, which has a duct-like behaviour \citep{towne2017acoustic}, and the evanescent $k_p^\pm$ waves with $n_r\geq 2$, which behave like modes in a soft duct as well at this frequency. In this regard, we note that the $k_p$ eigenvalues with $k_i>0$ are associated with $k^+$ waves, whereas the eigenvalues with $k_i<0$ are associated with $k^-$ modes. Within the vortex-sheet model, possible downstream-travelling transmitted modes are: (i) the K-H mode and its complex conjugate, (ii) the evanescent guided mode of first radial order with supersonic phase speed, (iii) the evanescent trapped modes of higher radial order with subsonic phase speed, and (iv) the acoustic modes. A summary of the modes involved in the reflection coefficient computation is reported in table \ref{tab:V-S_modes}.

\begin{table}
\centering
\begin{tabular}{cc}
\multicolumn{2}{c}{Vortex-sheet eigenmodes}\\
\hline
Incident & $k_{KH}^+$\\
\hline
\multirow{3}{*}{Reflected} & propagative $k_p^-$ with $n_r=2$\\
						   & evanescent $k_p^-$ with $n_r=1$\\
						   & propagative and evanescent $k_a^-$\\
\hline
\multirow{3}{*}{Transmitted} & $k_{KH}^+$ and $k_{KH}^{*+}$\\
                             & evanescent $k_p^+$ with $n_r\geq 1$\\
                             & propagative and evanescent $k_a^+$\\
\end{tabular}
\caption{Summary of the modes supported by the vortex-sheet model involved in the reflection-coefficient computation.}
\label{tab:V-S_modes}
\end{table}

\begin{figure}
\centering
\includegraphics[scale=0.27]{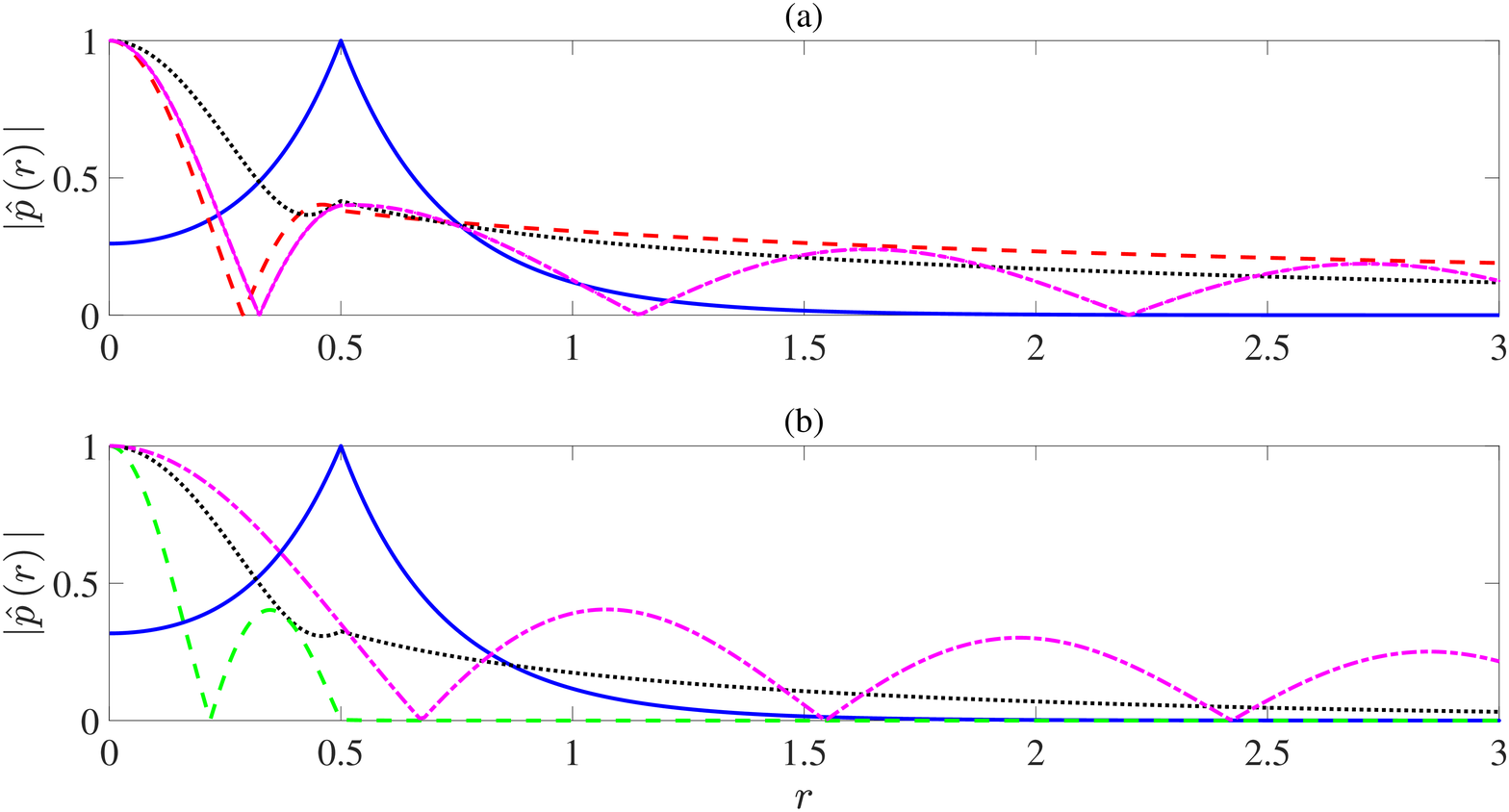}
\caption{Pressure eigenfunctions computed using the vortex-sheet model for $m=0$ and $St=0.68$. The colours are the same as those used in figures \ref{fig:eigenspectrum_VS_up} and \ref{fig:eigenspectrum_VS_down} to identify the different mode families. (a) incident and reflected waves upstream of the shock for $M_j=1.1$ and $T\approx 0.81$: solid blue line refers to the incident $k_{KH}^+$ wave, dashed red line to the propagative $k_p^-$ wave with $n_r=2$, dotted black line to the evanescent $k_p^-$ mode with $n_r=1$, dash-dotted magenta line to the propagative $k_a^-$ wave. (b) transmitted waves downstream of the shock for $M_{j_2}=0.91$ and $T\approx 0.85$: solid blue line refers to the $k_{KH}^+$ wave, dotted black line to the evanescent $k_p^+$ with $n_r=1$, dashed green line to the evanescent $k_p^+$ with $n_r=2$, dash-dotted magenta line to the propagative $k_a^+$ wave.}
\label{fig:eigenfunctions_VS}
\end{figure}

Examples of the normalised pressure eigenfunctions of the waves upstream and downstream of the shock are reported in figure \ref{fig:eigenfunctions_VS}. As expected, the incident and transmitted K-H waves show a peak at the vortex-sheet location and the reflected waves, that is $k_p^-$ and $k_a^-$ modes, have a support both inside and outside the jet. This behaviour is also detected downstream of the shock for the transmitted evanescent $k_p^+$ wave with $n_r=1$. On the other hand, the transmitted $k_p^+$ modes with $n_r>1$ show a spatial support concentrated inside the jet, consistent with their identity as acoustic waves trapped within the jet core.

\begin{figure}
\centering
\includegraphics[scale=0.27]{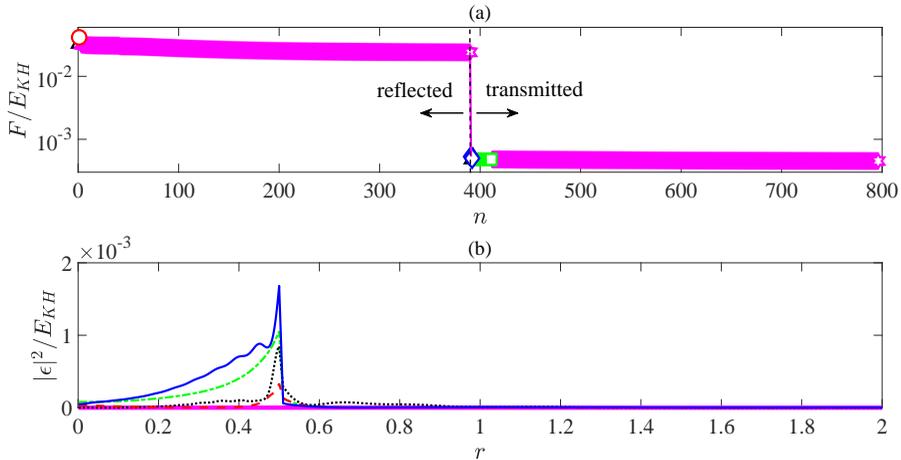}
\caption{(a) Evolution of the normalised objective function to minimise in the reflection-coefficient calculation as a function of the number of modes considered (markers and colours are the same as those used to identify the modes in the eigenspectra in figures \ref{fig:eigenspectrum_VS_up} and \ref{fig:eigenspectrum_VS_down}). (b) normalised amplitude of the error densities associated with each conservation equation as a function of the radial distance $r$ for $N=797$: solid blue line refers to the mass conservation equation, dashed red line to the $x$ momentum, dotted black line to the $r$ momentum, bold magenta line to the $\theta$ momentum, dash-dotted green line to the energy equation.}
\label{fig:errors_VS}
\end{figure}

The modes described above are used to minimise the error objective function $F$ in \eqref{eq:minimization} via the pseudo-inverse solution \eqref{eq:pinv}. Figure \ref{fig:errors_VS}a shows the evolution of the objective function $F$ normalised by the energy norm of the incident K-H wave as a function of the number of reflected and transmitted modes $n=n_R+n_T$ considered. We here make a convergence analysis showing how the objective function changes by adding reflected and transmitted modes. We choose to add modes as a function of the family they belong to. For each mode family, waves are added by increasing $\vert k\vert$. Specifically, we first add in the calculation the reflected waves, that is the $k_p^-$ and $k_a^-$ mode families, and then we add the transmitted modes downstream of the shock, that is $k_{KH}^+$ and its complex conjugate, the $k_p^+$ modes for $n_r\geq 1$ and the $k_a^+$ mode, for a total number of modes $N=N_R+N_T=797$. We note that the objective function remains approximately constant when reflected waves are added. When added, the transmitted K-H mode provides a significant decay in the cost function, which indicates that it has an important role in the sharp discontinuity. This is expected, as the impinging and transmitted K-H modes have a similar structure. The addition of the other transmitted modes has smaller impacts on the cost functional, which saturates at a value of  $\approx 4.5\cdot 10^{-4}$. The amplitude of the squared error densities associated with each conservation equation as a function of the radial distance $r$ for $N=797$ modes is shown in figure \ref{fig:errors_VS}b. Similar to the objective function, the error densities are normalised by the energy norm of the incident $k_{KH}$ wave. All of the error densities, except for the tangential momentum equation, which is 0 for all $r$ for $m=0$, exhibit a peak at the vortex-sheet location $r=0.5$, where a discontinuity in the mean-flow profiles is found. We also note that the error-density amplitude is larger for inner radial position but never exceeds $2\cdot 10^{-3}$.

\begin{figure}
\centering
\includegraphics[scale=0.27]{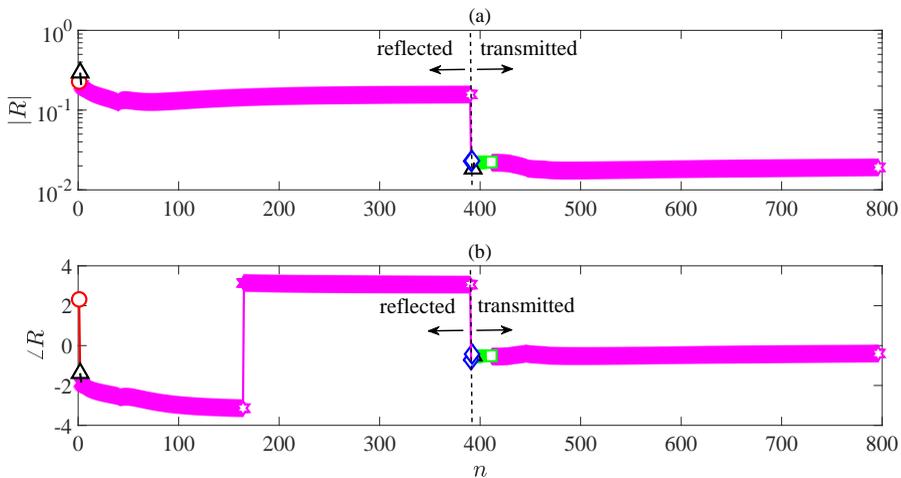}
\caption{Evolution of the reflection coefficient as a function of the modes considered in the calculation (markers and colours are the same as those used to identify the modes in the eigenspectra in figures \ref{fig:eigenspectrum_VS_up} and \ref{fig:eigenspectrum_VS_down}): (a) amplitude, (b) phase.}
\label{fig:R_VS}
\end{figure}

Figure \ref{fig:R_VS} shows the evolution of the amplitude and phase of the reflection coefficient associated with the upstream-travelling guided mode of second radial order as a function of the number of modes $n$. Similar to the objective function in figure \ref{fig:errors_VS}a, both the amplitude and phase of the reflection coefficient show a significant jump when modes on the transmitted side are included in the calculation. Specifically, the solution does not drastically change after the inclusion of the $k_{KH}^+$ mode and remains approximately constant. The solution converges to $1.9\cdot 10^{-2}$ and $-0.4$ for the amplitude and phase, respectively.

\subsection{Finite-thickness model}
\label{sec:LEE_results}
\begin{figure}
\centering
\includegraphics[scale=0.25]{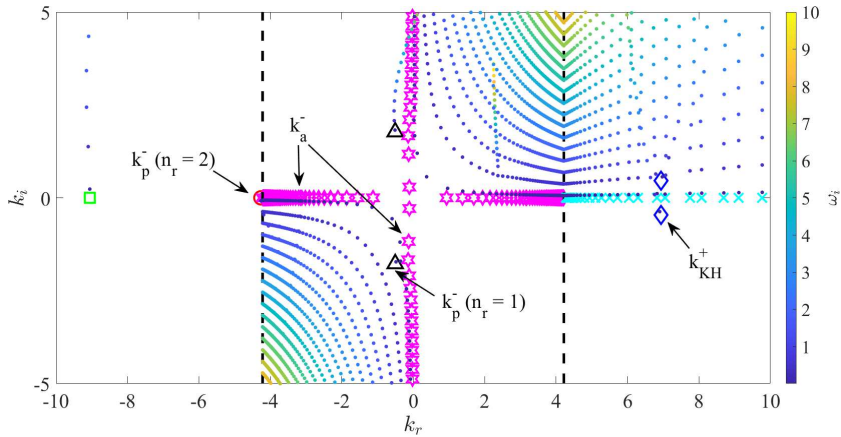}
\caption{Eigenspectrum upstream of the shock computed using the Rayleigh equation with a shear-layer thickness $R/\theta_R=10$ for azimuthal mode $m=0$, $T\approx 0.81$ and $M_{j_1}=1.1$. Empty markers represent eigenvalues for $\omega\in\mathcal{R}$, $\bullet$ eigenvalues for $\omega\in\mathcal{C}$, with $\omega_i\to+\infty$ from blue to yellow. Markers and colours are the same as those used for the VS in figure \ref{fig:eigenspectrum_VS_up}. The modes that are only supported by the finite-thickness model, that is the $k_{cr}^+$ modes, are here indicated by cyan $\times$. Dashed lines refer to the sonic speed $\pm c_\infty$.}
\label{fig:eigenspectrum_RE_up}
\end{figure}

Figure \ref{fig:eigenspectrum_RE_up} shows the eigenspectrum upstream of the shock computed using linear stability theory for a shear layer with finite thickness $R/\theta_R=10$. Similar to the vortex-sheet model, we find $k_{KH}^+$ and $k_{KH}^{*+}$ waves, the upstream-travelling propagative $k_p^-$ wave with $n_r=2$, the evanescent $k_p^\pm$ waves with $n_r=1$, the propagative $k_p^+$ modes for $n_r>1$ and the propagative and evanescent $k_a^\pm$ modes. Additionally, the finite-thickness model supports one family of waves that are not supported by the vortex sheet: critical-layer modes, denoted hereinafter $k_{cr}$. These modes have positive, subsonic phase and group velocities and lie on the real axis. Their spatial support is concentrated in the critical layer of the jet, i.e., the region of the jet where the phase speed equals the local mean flow velocity \citep{tissot2017sensitivity}. Critical-layer modes with small wavenumbers are characterised by a spatial support mainly concentrated in the core of the jet and possess a phase speed close to the mean jet velocity, whereas $k_{cr}$ modes with larger wavenumbers are mostly concentrated in the shear layer and have a phase speed value which decreases as the spatial support of the mode moves more and more outside of the jet. To summarise, similar to the V-S model, the finite-thickness model supports two families of reflected waves upstream of the shock: (i) the propagative and evanescent guided modes $k_p^-$ with $n_r=2$ and $1$, respectively, and (ii) the propagative and evanescent free-stream acoustic waves $k_a^-$.

\begin{figure}
\centering
\subfigure[]{\includegraphics[scale=0.25]{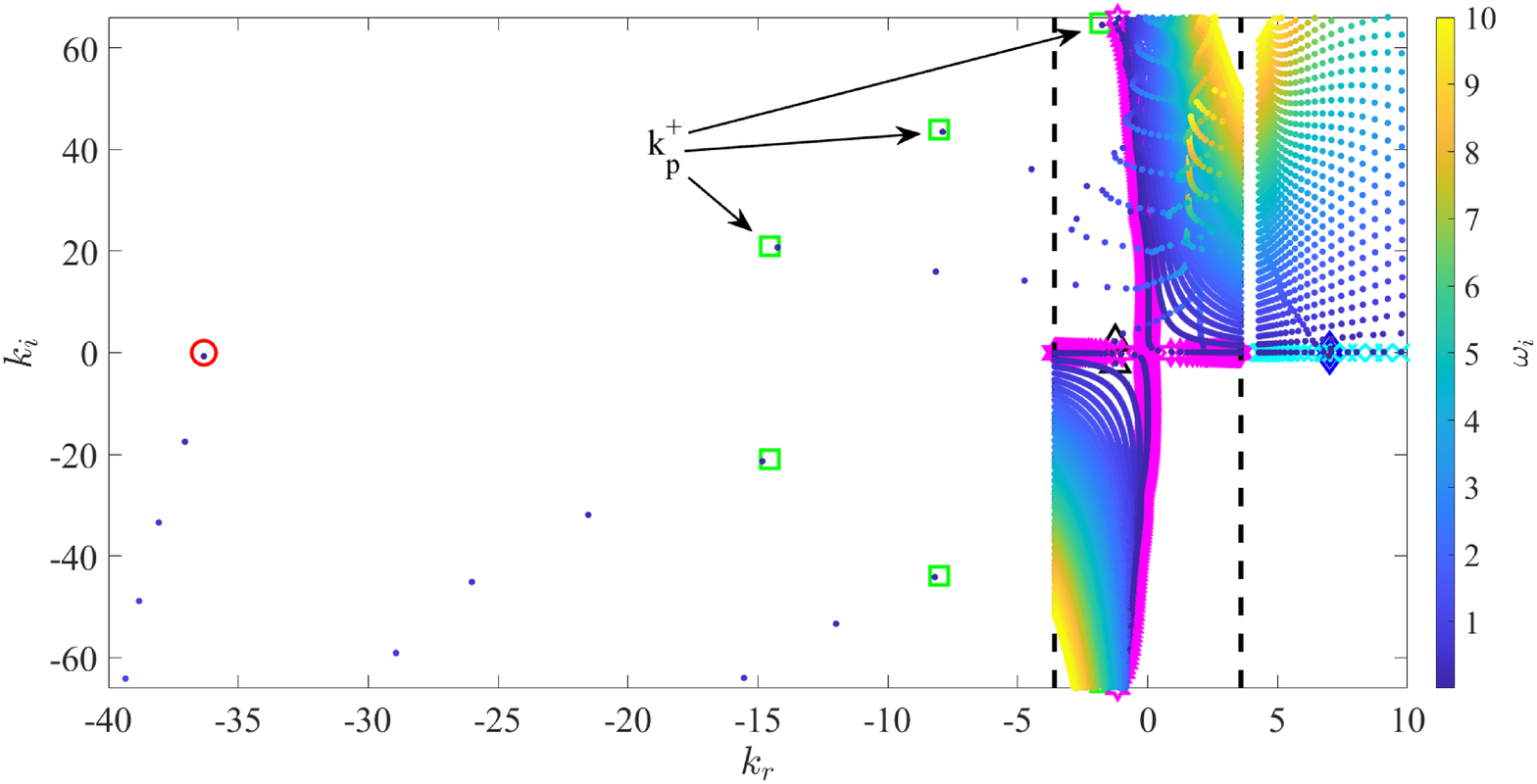}}
\subfigure[]{\includegraphics[scale=0.25]{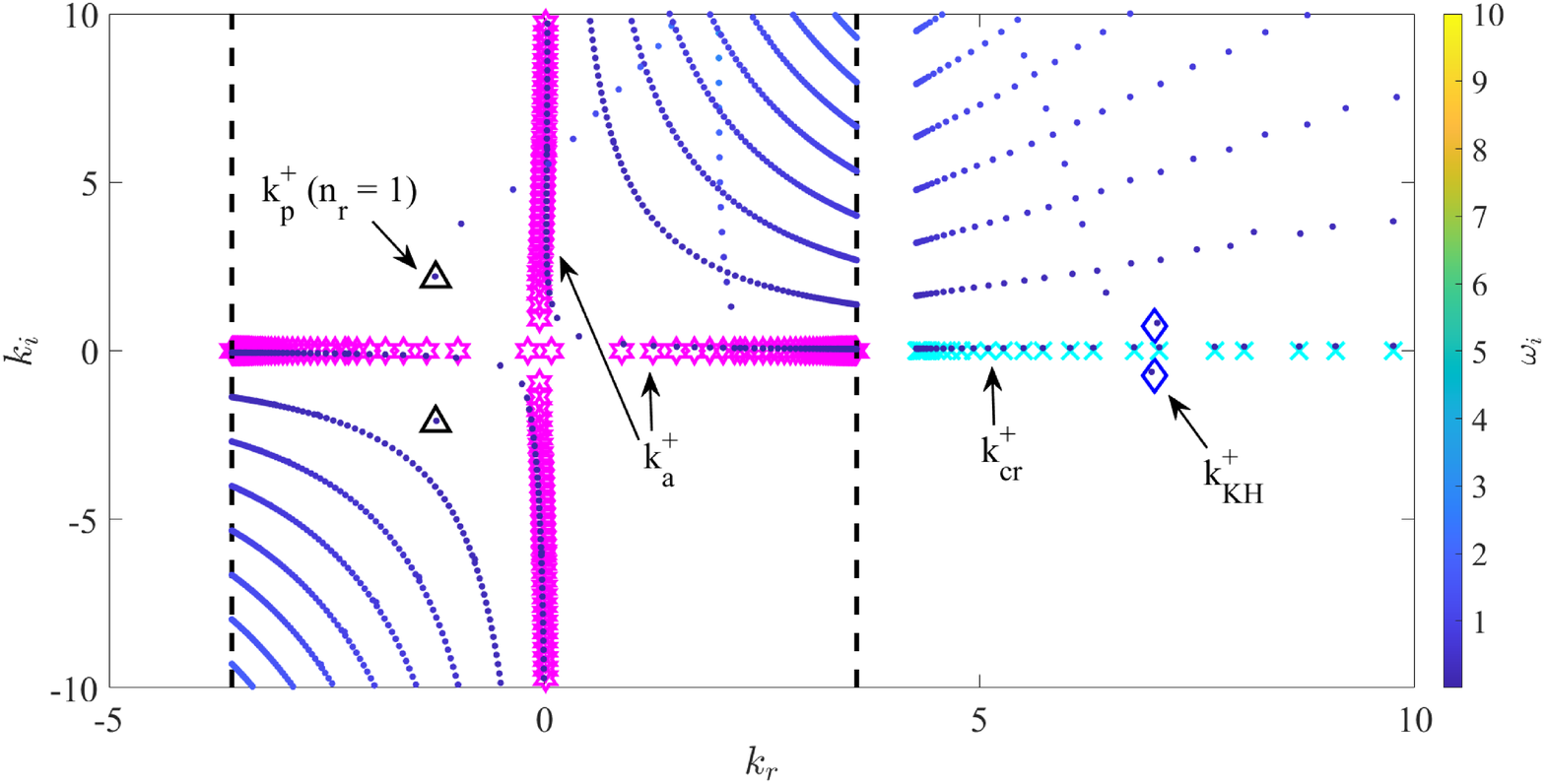}}
\caption{Eigenspectrum downstream of the shock for azimuthal mode $m=0$, $T\approx 0.85$ and $M_{j_2}=0.91$ and $M_{a_2}=0.84$: (a) eigenspectrum, (b) zoom around the origin. Empty markers represent eigenvalues from vortex-sheet model for $\omega\in\mathcal{R}$, $\bullet$ for $\omega\in\mathcal{C}$,with $\omega_i\to+\infty$ from blue to yellow. Colours and markers are the same as those used for the VS in figure \ref{fig:eigenspectrum_VS_down}. The modes that are only supported by the finite-thickness model, that is the $k_{cr}^+$ modes, are here indicated by cyan $\times$.}
\label{fig:eigenspectrum_RE_down}
\end{figure}

Figure \ref{fig:eigenspectrum_RE_down} shows the eigenspectrum downstream of the shock computed using a finite-thickness model. Consistent with the vortex sheet, we find the $k_{KH}^+$ and $k_{KH}^{*+}$ waves, the evanescent $k_p^\pm$ waves with $n_r=1$ with negative supersonic phase speed, the propagative $k_d^-$ wave for $n_r=2$ and the evanescent $k_p^\pm$ with $n_r\geq2$. Unlike those computed by the vortex sheet, the evanescent guided modes for $n_r>1$ bend towards the supersonic phase speed region as $n_r$ increases and eventually merge with the evanescent free-stream acoustic modes for $n_r>4$. Additionally, we observe the $k_{cr}^+$ waves that are not supported by the VS. To summarise, the transmitted modes we use for the reflection-coefficient computation are: (i) the K-H wave and its complex conjugate, (ii) the evanescent $k_p^+$ wave with $n_r=1$ with supersonic phase speed, (iii) the evanescent $k_p^+$ waves with $n_r>1$ with subsonic phase speed, (iv) the propagative and evanescent $k_a^+$ waves, and (v) the critical-layer modes. Unlike the computation performed using the V-S model and consistent with the eigenspectrum shown in figure \ref{fig:eigenspectrum_RE_down}, we use evanescent $k_p^+$ waves on the transmitted side up to the radial order $n_r=4$, that is before the eigenvalues merge with the evanescent $k_a^+$ modes and become no longer distinguishable. A summary of all the waves involved in the calculation process for the evaluation of $R$ is reported in table \ref{tab:LEE_modes}.

\begin{table}
\centering
\begin{tabular}{cc}
\multicolumn{2}{c}{Eigenmodes from finite-thickness model}\\
\hline
Incident & $k_{KH}^+$\\
\hline
\multirow{3}{*}{Reflected} & propagative $k_p^-$ with $n_r=2$\\
						   & evanescent $k_p^-$ with $n_r=1$\\
						   & propagative and evanescent $k_a^-$ waves\\
\hline
\multirow{4}{*}{Transmitted} & $k_{KH}^+$ and $k_{KH}^{*+}$\\
                             & evanescent $k_p^+$ with $n_r\geq 1$\\
                             & propagative and evanescent $k_a^+$ waves\\
						     & $k_{cr}^+$ waves
\end{tabular}
\caption{Summary of the modes supported by the finite-thickness model used to compute the reflection coefficient.}
\label{tab:LEE_modes}
\end{table}

\begin{figure}
\centering
\includegraphics[scale=0.27]{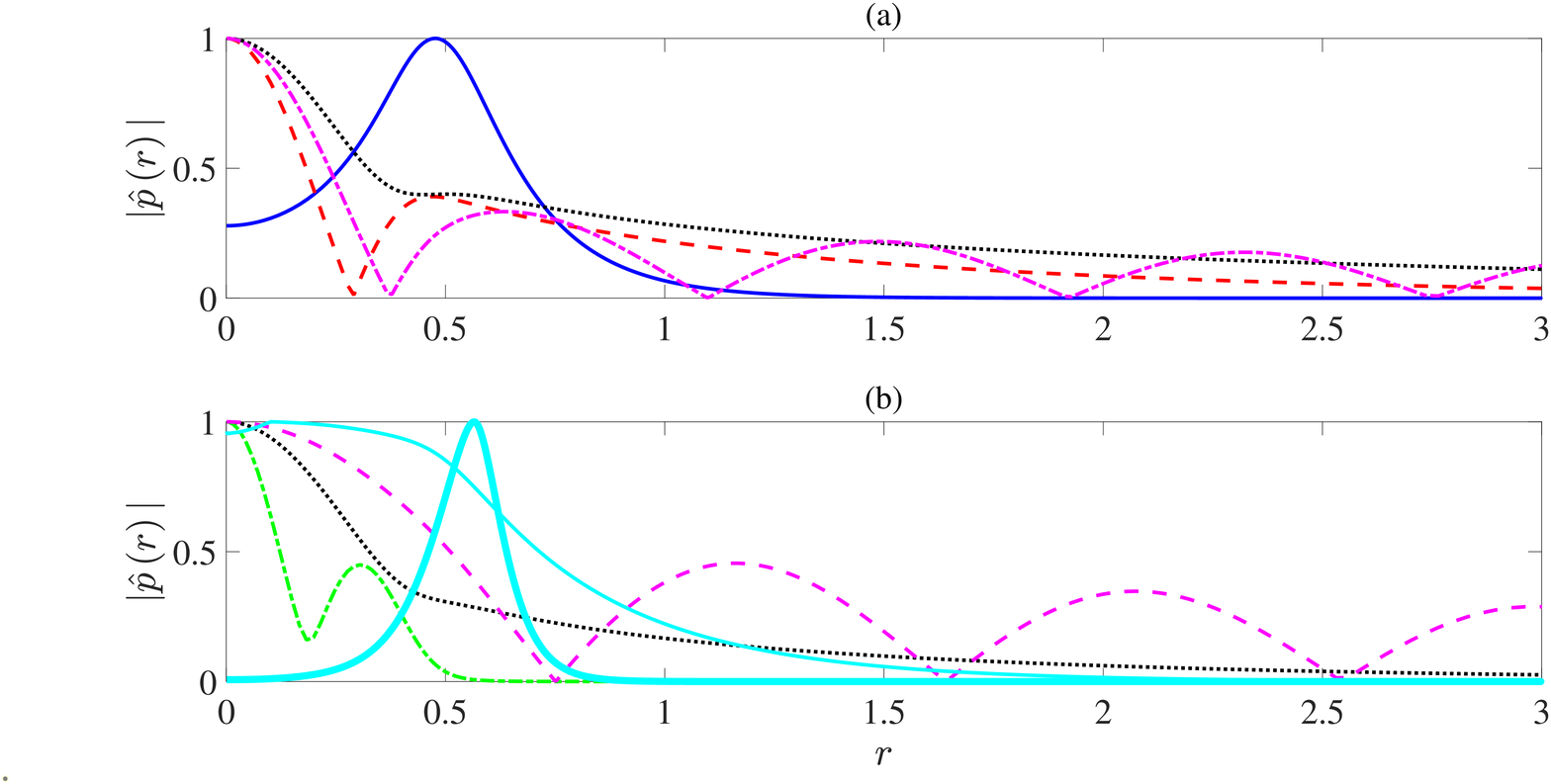}
\caption{Pressure eigenfunctions computed using the finite-thickness model for $m=0$ and $St=0.68$. The colours are the same as those used in figures \ref{fig:eigenspectrum_RE_up} and \ref{fig:eigenspectrum_RE_down} to identify the different mode families. (a) incident and reflected waves upstream of the shock for $M_j=1.1$ and $T\approx 0.81$: solid blue line refers to $k_{KH^+}$ wave, dashed red line to $k_p^-$ mode with $n_r=2$, dotted black line to $k_p^-$ mode with $n_r=1$, magenta dash-dotted line to $k_a^-$ wave. (b) transmitted waves downstream of the shock for $M_j=0.91$ and $T\approx 0.85$: dotted black line refers to $k_p^+$ mode with $n_r=1$, dash-dotted green line to $k_p^+$ mode with $n_r=2$, dashed magenta line to $k_a^+$ wave, solid and bold cyan lines to $k_{cr}^+$ modes with $k\approx 4.28$ and $18.88$, respectively.}
\label{fig:eigenfunctions_RE}
\end{figure}

Examples of the pressure eigenfunctions of the waves upstream and downstream of the shock computed using the finite-thickness model are shown in figure \ref{fig:eigenfunctions_RE}. Similar to the VS, the incident K-H mode shows an eigenfunction with a peak located in the centre of the shear layer and the reflected $k_p^-$ and $k_a^-$ waves are characterised by a peak on the centreline and have a spatial support both inside and outside of the jet. The eigenfunctions of the $k_p^+$ and $k_a^+$ waves have a similar shape to that found using the V-S model downstream of the shock. As mentioned above, $k_{cr}^+$ modes have eigenfunctions with a spatial support concentrated either inside of the jet or in the shear layer depending on the wavenumber value considered. The spatial support of the critical-layer modes moves to larger $r$ and the phase velocity $U_{\phi}$ decreases as $\vert k_{cr}\vert$ increases.

\begin{figure}
\centering
\includegraphics[scale=0.28]{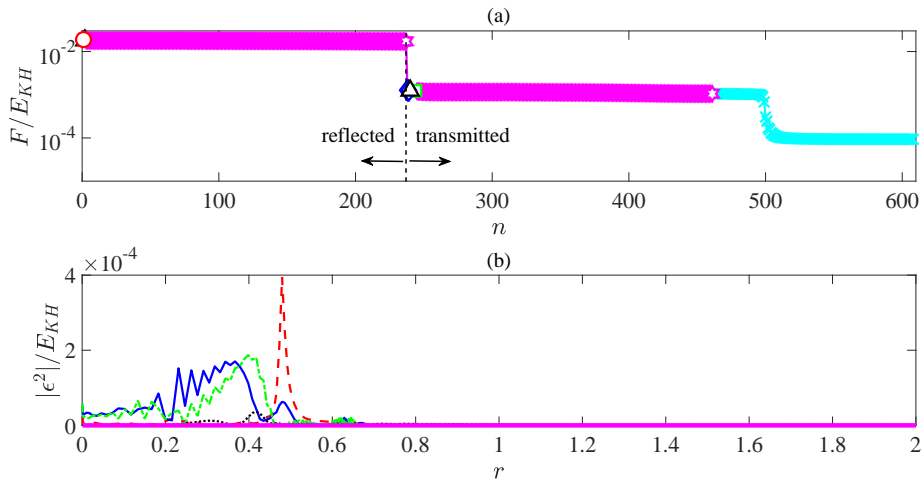}
\caption{(a) Evolution of the normalised objective function to minimise in the reflection-coefficient calculation as a function of the number of modes considered. Markers and colours are the same as those used in figures \ref{fig:eigenspectrum_RE_up} and \ref{fig:eigenspectrum_RE_down} to identify the mode families. (b) Normalised amplitude of the error densities associated with each conservation equation as a function of the radial distance $r$ for $N=608$: solid blue line refers to the mass conservation equation, dashed red line to the $x$ momentum, dotted black line to the $r$ momentum, bold magenta line to the $\theta$ momentum, dash-dotted green line to the energy equation.}
\label{fig:errors_RE}
\end{figure}

Figure \ref{fig:errors_RE}a shows the evolution of the objective function $F$ as a function of the number of modes $n$. A behaviour similar to that observed using the VS is detected when modes that are supported by both the V-S and finite-thickness models are considered. When critical-layer modes are added in the computation, we observe an important reduction of the cost function, which seems to saturate at a value of $9\cdot 10^{-5}$. Figure \ref{fig:errors_RE}b shows the trend of the amplitude of the squared error densities as a function of the radial distance for $N=608$. We note that the error densities are larger in the inner part of the jet and they become more significant near the shear layer. Specifically, the largest error occurs for the momentum equation along the axial direction, with an error density peak located in the centre of the shear layer ($r\approx 0.5$); however, the value of the error density never exceeds $4\cdot 10^{-4}$.

\begin{figure}
\centering
\includegraphics[scale=0.28]{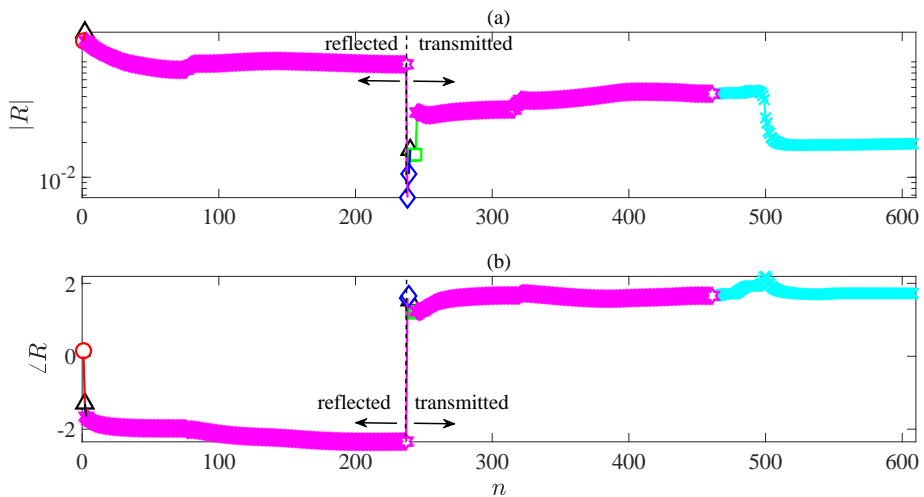}
\caption{Evolution of the amplitude and phase of the reflection coefficient as a function of the number of modes considered. Markers and colours are the same as those used in figures \ref{fig:eigenspectrum_RE_up} and \ref{fig:eigenspectrum_RE_down} to identify the mode families. (a) amplitude, (b) phase.}
\label{fig:R_RE}
\end{figure}

\begin{figure}
\centering
\includegraphics[scale=0.27]{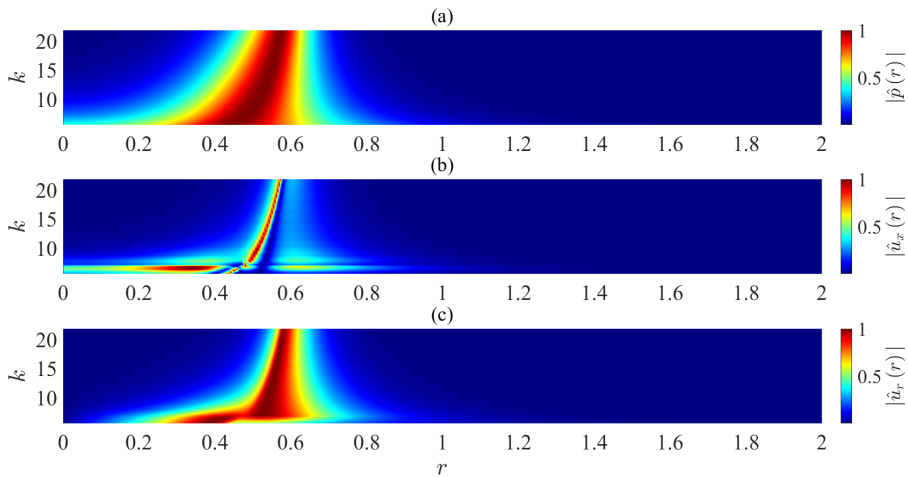}
\caption{Normalised eigenfunctions of $k_{cr}^+$ modes computed using the finite-thickness model as a function of wavenumber $k$ and radial distance $r$: (a) pressure, (b) axial velocity, (c) radial velocity.}
\label{fig:eigenfunctions_cr}
\end{figure}

The evolution of the amplitude and phase of the reflection coefficient as a function of $n$ is shown in figure \ref{fig:R_RE}. Similar to the V-S model, we see that a significant change in the value of both amplitude and phase occurs when modes on the transmitted side are included in the calculation. Consistent with what we observed in the objective function trend, the addition of the $k_{cr}^+$ modes appears to play an important role in the determination of the reflection coefficient. Specifically, a significant amplitude drop occurs for critical-layer modes with $5.5\leq \vert k\vert\leq 21$, after which both the amplitude and phase values remain approximately constant. These are modes with wavelength in the range $\approx\left[0.3D,D\right]$ and, hence, 6 to 20 times the shear-layer thickness. Their phase speed is $\approx$ $0.2\,U_j\leq U_{\phi}\leq 0.75\,U_j$ and they are mostly concentrated in the centre of the shear layer, as shown in figure \ref{fig:eigenfunctions_cr}.

\subsubsection{Comparison with vortex-sheet results}
\label{subsubsec:VS_LEE_comparison}
\begin{figure}
\centering
\includegraphics[scale=0.27]{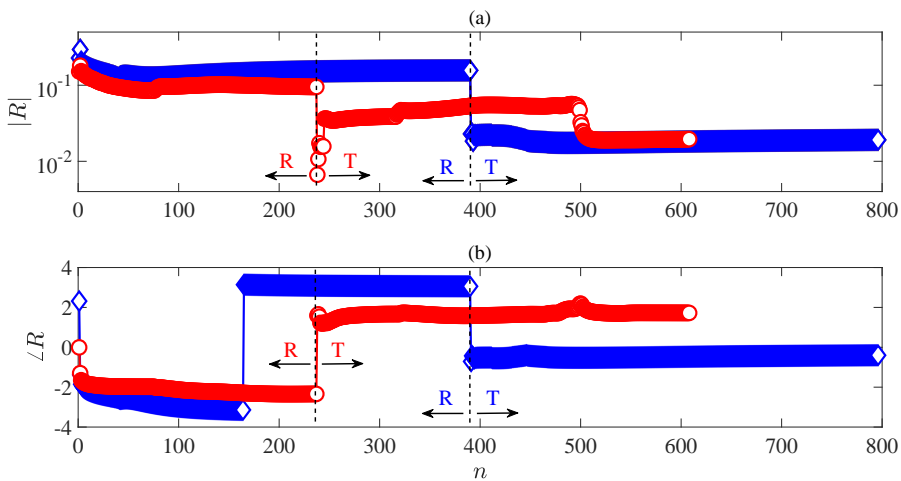}
\caption{Comparison of the evolution of the reflection-coefficient amplitude and phase values as a function of the number of modes $n$ obtained using the vortex-sheet and finite-thickness models: blue $\diamond$ refer to VS, red $\circ$ to finite-thickness (FT) model. (a) amplitude, (b) phase.}
\label{fig:R_VS_RE}
\end{figure}

Figure \ref{fig:R_VS_RE} shows the evolution of the amplitude and phase values of the reflection coefficient as a function of $n$ obtained using both the vortex sheet and finite-thickness models. Despite its simplicity and the impossibility of describing shear-layer dynamics, the V-S model provides the same value for the reflection-coefficient amplitude as that obtained using the finite-thickness model. On the contrary, a larger discrepancy between the two flow models is found for the phase value and this discrepancy could be likely ascribed to shear-layer thickness effects.  A summary of the reflection-coefficient values obtained for both the flow models is reported in table \ref{tab:R_VS_RE}.

\begin{table}
\centering
\begin{tabular}{ccc}
\multicolumn{3}{c}{Reflection-coefficient values}\\
\hline
 & Amplitude & Phase\\
VS & 0.019 & -0.4\\
FTM & 0.0195 & 1.7\\
\end{tabular}
\caption{Summary of the reflection-coefficient values obtained using the vortex-sheet and finite-thickness model to describe the jet dynamics.}
\label{tab:R_VS_RE}
\end{table}

\begin{figure}
\centering
\includegraphics[scale=0.27]{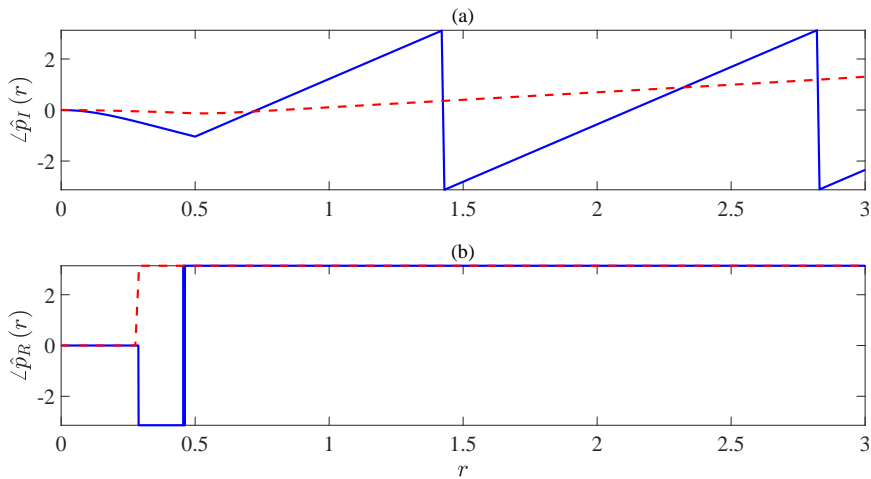}
\caption{Phase of the incident K-H and reflected guided mode pressure eigenfunctions for both flow models: blue solid lines represent the vortex sheet, dashed red lines the finite-thickness model. (a) incident $k_{KH}^+$ wave, (b) reflected $k_p^-$ mode with $n_r=2$.}
\label{fig:phase_p}
\end{figure}

To explore in more detail the phase discrepancy provided by the two flow models, the phase of the incident $k_{KH}^+$ and reflected $k_p^-$ modes for both the V-S and finite-thickness models is reported in figure \ref{fig:phase_p}. We report results for $r\leq 3$ given that the phase of the reflected wave does not change and the amplitude of the incident K-H wave is zero for larger radial distances. The phase of the pressure eigenfunction of the reflected wave is zero for $r\leq 0.3$ and goes to $\pm\pi$ for larger radial distances with no differences between the two flow models. On the contrary, a significant discrepancy between the results provided by the two flow models is observed for the incident K-H wave. For the V-S model, the phase decreases to $-1$ for $r\approx 0.5$ and then goes quite sharply to $\pi$ and oscillates between $\pi$ and $-\pi$ for all the radial distances. On the other hand, for the finite-thickness model, the phase of the incident pressure eigenfunction remains close to zero in the inner region of the jet and then gradually increases and eventually goes to $\pi$ for very large radial distances. Such different radial evolution of the phase of the incident pressure eigenfunction could explain the discrepancy observed in the phase of the reflection coefficient obtained from the V-S and finite-thickness models. Further analyses should be carried out to fully understand this behaviour, but this is beyond the scope of the present manuscript.

\subsubsection{Reflected and transmitted fields}
\label{subsubsec:R_T_fields}
The reflected and transmitted pressure fields for the flow condition investigated above using a finite-thickness model are here reconstructed in the $x$-$r$ plane. Using the normal-mode ansatz \eqref{eq:normal_ansatz}, the reflected and transmitted fields are given by

\begin{subequations}
\begin{equation}
p_R\left(x,r,\theta\right)=\sum_{n_R=1}^{N_R}R_{n_R}\hat{p}_{1R,n_R}e^{i\left(k_{nR}x+m\theta\right)}\mathrm{,}
\label{eq:p_R}
\end{equation}
\begin{equation}
p_T\left(x,r,\theta\right)=\sum_{n_T=1}^{N_T}T_{n_T}\hat{p}_{2T,n_T}e^{i\left(k_{nT}x+m\theta\right)}\mathrm{.}
\label{eq:p_T}
\end{equation}
\end{subequations}

\begin{figure}
\centering
\includegraphics[scale=0.28]{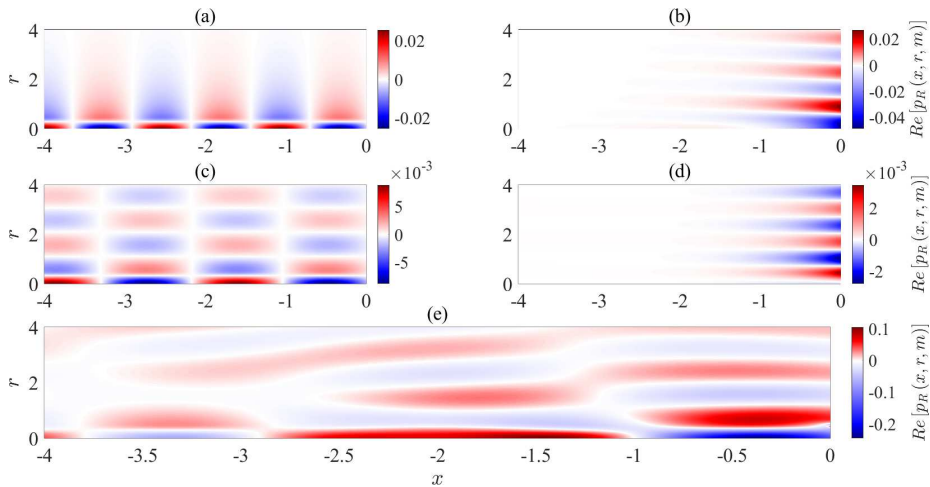}
\caption{Reconstruction of the reflected fields in the $x$-$r$ plane for a shock discontinuity located at $x=0$: (a) $k_p^-$ mode with $n_r=2$, (b) $k_p^-$ mode with $n_r=1$, (c) propagative $k_a^-$ wave, (d) evanescent $k_a^-$ wave, (e) total field.}
\label{fig:p_field_R}
\end{figure}

We set $x=0$ to be the location of the shock discontinuity, so the reflected upstream-travelling and transmitted downstream-travelling waves evolve along negative and positive $x$ directions, respectively.  Figure \ref{fig:p_field_R} shows the entire reflected field as well as the reflected fields obtained using for the reconstruction each type of the upstream-travelling modes considered in the reflection-coefficient calculation, that is the propagative $k_p^-$ mode with $n_r=2$, the evanescent $k_p^-$ mode with $n_r=1$ and an example of the propagative and evanescent $k_a^-$ waves with $\vert k\vert\approx 2.79$ and $2.1$, respectively. As expected, the fields reconstructed using the $k_p^-$ modes show a spatial support both inside and outside the jet with a prescribed radial decay. Regarding the axial evolution, the evanescent $k_p^-$ mode with $n_r=1$ exhibits a decay starting from the shock position $x=0$, whereas the $k_p^-$ mode with $n_r=2$ has a fixed axial structure consistent with its neutrally-stable nature. A similar behaviour is found for the pressure fields reconstructed using propagative and evanescent free-stream acoustic waves. The entire reflected pressure field, which results from the linear superposition of all the reflected $k^-$ waves, shows an axially and radially decaying structure starting from $x=0$ and $r=0$, respectively.

\begin{figure}
\centering
\includegraphics[scale=0.28]{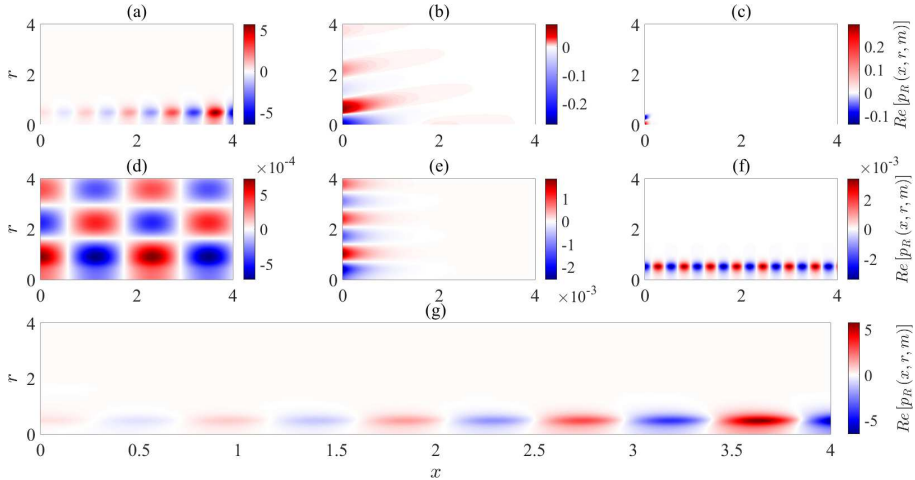}
\caption{Reconstruction of the transmitted fields in the $x$-$r$ plane for a shock discontinuity located at $x=0$: (a) $k_{KH}^+$ wave, (b) supersonic $k_p^-$ mode with $n_r=1$, (c) subsonic $k_p^+$ with $n_r=2$ (d) propagative $k_a^+$ wave, (e) evanescent $k_a^+$ wave, (f) $k_{cr}^+$ mode, (g) total field.}
\label{fig:p_field_T}
\end{figure}

The entire transmitted field as well as the transmitted fields obtained using each mode family considered in the reflection-coefficient calculation are shown in figure \ref{fig:p_field_T}. We here report the fields obtained using the $k_{KH}^+$ wave, the evanescent supersonic $k_p^+$ mode with $n_r=1$, the evanescent subsonic $k_p^+$ mode with $n_r=2$, the propagative and evanescent $k_a^+$ waves with $\vert k\vert\approx 2.7$ and $2.8$, respectively, and the $k_{cr}^+$ mode for $\vert k\vert\approx 11.5$. As expected, the shape and intensity of the total transmitted field is dominated by the unstable K-H mode. The shape of the transmitted fields reconstructed using the individual mode family is consistent with the neutral/evanescent nature and radial structure described above.

\subsubsection{Frequency-Mach-number dependence of the reflection coefficient}
\label{subsubsec:R_StM}
\begin{figure}
\centering
\includegraphics[scale=0.28]{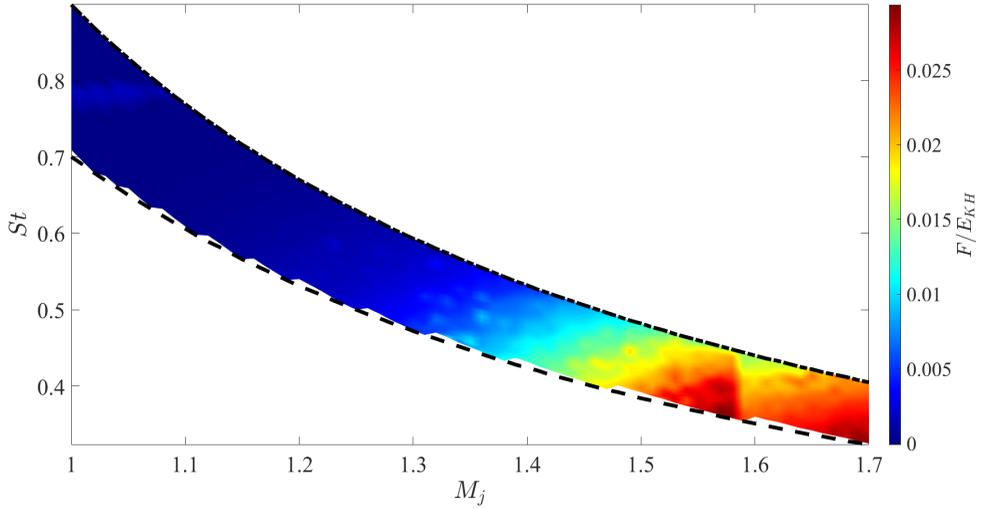}
\caption{Normalised error objective function as a function of Strouhal and jet Mach numbers for a finite-thickness model. Dashed and dash-dotted lines refer to the branch- and saddle-point locations, respectively.}
\label{fig:F_MSt}
\end{figure}

\begin{figure}
\centering
\includegraphics[scale=0.28]{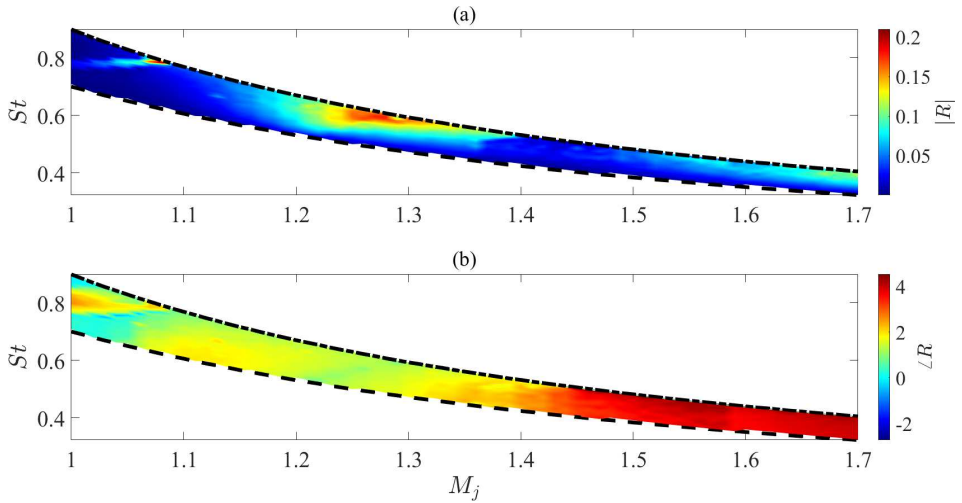}
\caption{Evolution of the reflection coefficient as a function of Strouhal and jet Mach numbers for a finite-thickness model. Dashed and dash-dotted lines refer to the branch- and saddle-point locations, respectively. (a) amplitude, (b) phase.}
\label{fig:R_MSt}
\end{figure}

We here explore the frequency-Mach-number dependence of the reflection coefficient associated with the upstream-travelling guided mode with $n_r=2$. For this purpose, we let the jet Mach number vary in the range $M_j=\left[1,1.7\right]$ and explore the Strouhal-number band for which the guided mode is propagative for such flow conditions, that is the $St$-number range delimited by the branch and saddle points (see \cite{mancinelli2019screech} and \cite{mancinelli2021complexvalued}). The resolution for both $\Delta M_j$ and $\Delta St$ was set equal to $10^{-2}$. Figure \ref{fig:F_MSt} shows the evolution of the objective function $F$ normalised by the energy norm of the incident K-H mode as a function of $M_j$ and $St$. The errors are very small for low $M_j$, five orders of magnitude lower than the energy norm of the incident K-H mode, and gradually rise as the jet Mach number increases. This behaviour is likely related to reduced efficiency of the mode-matching approach for the stronger discontinuities associated with increasing $M_j$. Nevertheless, the maximum value of the normalised error never exceeds 3\% of the incident energy. The amplitude and unwrapped phase of the reflection coefficient as a function of $St$ and $M_j$ are shown in figure \ref{fig:R_MSt}. The reflection-coefficient amplitude rises with increasing $M_j$ for a given $St$ and appears to be larger in the proximity of the saddle point for all $M_j$, with the jet-flow region $M_j=\left[1.17,1.37\right]$ showing the highest amplitude. Overall, the phase gradually increases for larger $M_j$ and lower $St$.

\section{Conclusions}
\label{sec:conclusions}
The scattering of a Kelvin-Helmholtz wave into a discrete, upstream-travelling guided mode due to a normal shock in a jet flow was computed in the present manuscript. Jets with zero- and finite-thickness shear layers were modelled with an analytical vortex sheet and a numerical solution of the Rayleigh equation, respectively. The reflection coefficient associated with the scattered, upstream-travelling wave was calculated via a mode-matching technique enforcing conservation equations through the shock discontinuity. Assuming small disturbances, the shock equations were linearised about the mean flow and the possible scattered modes, i.e., upstream-travelling reflected waves and downstream-travelling transmitted waves, were identified by assessing the corresponding group velocity using the Briggs-Bers criteria. Solutions based on a truncated set of modes were obtained via the minimisation of the error in the conservation equations, which corresponds to a weighted pseudo-inverse solution. The finite-thickness model showed that critical-layer modes, characterised by eigenfunctions with radial support mostly concentrated in the centre of the shear layer, by a wavelength 6 to 20 times the shear-layer thickness and by a phase velocity between 0.2 and 0.75 of the jet velocity, play an important role in the reflection-coefficient calculation and, thus, most likely in the physical mechanisms underpinning the process. Even though it is incapable of describing shear-layer dynamics, the simplified vortex-sheet model predicts the same value of the reflection-coefficient amplitude as the finite-thickness model. A larger discrepancy was found for the reflection-coefficient phase. The reconstructed reflected and transmitted pressure fields in the $x$-$r$ plane showed an organised structure consistent with the axial and radial evolution of the modes calculated by the finite-thickness model. The frequency-Mach-number dependence of the reflection coefficient has been explored revealing that the reflection-coefficient amplitude is larger for frequencies in the proximity of the saddle point and for jet Mach number values in between 1.17 and 1.37. The phase exhibited a gradual increase as the Strouhal and jet Mach numbers decreased and increased, respectively.

The mode-matching approach described in this paper appears to be a relevant and promising tool for better describing and predicting resonant dynamics found in jets, such as screech in supersonic jets. A similar approach may be developed to evaluate the reflection coefficient at the nozzle exit, thus providing all the elements to make screech-frequency predictions without any input from data using the model presented in \cite{mancinelli2021complexvalued}.

\appendix

\section{Linearised Euler Equations}
\label{sec:Euler_app}
The non-dimensional Euler equations in cylindrical coordinates are

\begin{subequations}
\begin{empheq}[left=\empheqlbrace]{align}
& \dfrac{D\rho}{Dt}+\rho\nabla\cdot\vec{u}=0\mathrm{,}\\
& \rho\dfrac{Du_x}{Dt}=-\dfrac{\partial p}{\partial x}\mathrm{,}\\
& \rho\left(\dfrac{Du_r}{Dt}-\dfrac{u_\theta^2}{r}\right)=-\dfrac{\partial p}{\partial r}\label{eq:momentum_r}\mathrm{,}\\
& \rho\left(\dfrac{Du_\theta}{Dt}+\dfrac{u_ru_\theta}{r}\right)=-\dfrac{1}{r}\dfrac{\partial p}{\partial \theta}\mathrm{,}\\
& \dfrac{DT}{Dt}+\left(\gamma -1\right)T\nabla\cdot\vec{u}=0\mathrm{,}\\
& p=\dfrac{\gamma - 1}{\gamma}\rho T\mathrm{,}
\end{empheq}
\label{eq:Euler}
\end{subequations}

\noindent where

\begin{subequations}
\begin{equation}
\dfrac{D}{Dt}=\dfrac{\partial}{\partial t}+u_x\dfrac{\partial}{\partial x}+u_r\dfrac{\partial}{\partial r}+\dfrac{u_\theta}{r}\dfrac{\partial}{\partial\theta}\mathrm{,}
\end{equation}
\begin{equation}
\nabla\cdot\vec{u}=\dfrac{\partial u_x}{\partial x}+\dfrac{\partial u_r}{\partial r}+\dfrac{u_r}{r}+\dfrac{1}{r}\dfrac{\partial u_\theta}{\partial \theta}\mathrm{.}
\end{equation}
\end{subequations}

Inserting the Reynolds decomposition \eqref{eq:Re_decomposition}, removing the mean and linearising, the LEE are written

\begin{subequations}
\begin{empheq}[left=\empheqlbrace]{align}
& \dfrac{\partial\rho}{\partial t}+\overline{u}_x\dfrac{\partial\rho}{\partial x}+u_r\dfrac{\partial\overline{\rho}}{\partial r}+\overline{\rho}\left(\dfrac{\partial u_x}{\partial x}+\dfrac{\partial u_r}{\partial r}+\dfrac{u_r}{r}+\dfrac{1}{r}\dfrac{\partial u_\theta}{\partial\theta}\right)=0\mathrm{,}\\
& \overline{\rho}\left(\dfrac{\partial u_x}{\partial t}+\overline{u}_x\dfrac{\partial u_x}{\partial x}+u_r\dfrac{\partial\overline{u}_x}{\partial r}\right)=-\dfrac{\partial p}{\partial x}\mathrm{,}\\
& \overline{\rho}\left(\dfrac{\partial u_r}{\partial t}+\overline{u}_x\dfrac{\partial u_r}{\partial x}\right)=-\dfrac{\partial p}{\partial r}\mathrm{,}\\
& \overline{\rho}\left(\dfrac{\partial u_\theta}{\partial t}+\overline{u}_x\dfrac{\partial u_\theta}{\partial x}\right)=-\dfrac{1}{r}\dfrac{\partial p}{\partial \theta}\mathrm{,}\\
& \dfrac{\partial T}{\partial t}+\overline{u}_x\dfrac{\partial T}{\partial x}+u_r\dfrac{\partial \overline{T}}{\partial r}+\left(\gamma -1\right)\overline{T}\left(\dfrac{\partial u_x}{\partial x}+\dfrac{\partial u_r}{\partial r}+\dfrac{u_r}{r}+\dfrac{1}{r}\dfrac{\partial u_\theta}{\partial\theta}\right)=0\mathrm{,}\\
& p=\dfrac{\gamma -1}{\gamma}\left(\overline{\rho}T+\overline{T}\rho\right)\mathrm{,}
\end{empheq}
\label{eq:LEE}
\end{subequations}

\noindent where we removed the primes from the fluctuating variables for notational simplicity. The locally parallel-flow assumption implies that the derivatives along the axial and azimuthal directions $x$ and $\theta$ are zero as well as the mean radial and azimuthal velocities $\overline{u}_r$ and $\overline{u}_\theta$. Assuming the normal mode ansatz \eqref{eq:normal_ansatz} yields

\begin{subequations}
\begin{empheq}[left=\empheqlbrace]{align}
& -i\omega\hat{\rho}+\overline{u}_xik\hat{\rho}+\dfrac{\partial\overline{\rho}}{\partial r}\hat{u}_r+\overline{\rho}\left(ik\hat{u}_x+\dfrac{\partial\hat{u}_r}{\partial r}+\dfrac{\hat{u}_r}{r}+\dfrac{im}{r}\hat{u_\theta}\right)=0\mathrm{,}\\
& \overline{\rho}\left(-i\omega\hat{u}_x+\overline{u}_xik\hat{u}_x+\dfrac{\partial\overline{u}_x}{\partial r}\hat{u}_r\right)=-ik\hat{p}\mathrm{,}\\
& \overline{\rho}\left(-i\omega\hat{u}_r+\overline{u}_xik\hat{u}_r\right)=-\dfrac{\partial\hat{p}}{\partial r}\mathrm{,}\\
& \overline{\rho}\left(-i\omega\hat{u}_\theta+\overline{u}_xik\hat{u}_\theta\right)=-\dfrac{im}{r}\hat{p}\mathrm{,}\\
& -i\omega\hat{T}_+\overline{u}_xik\hat{T}+\dfrac{\partial\overline{T}}{\partial r}\hat{u}_r+\left(\gamma -1\right)\overline{T}\left(ik\hat{u}_x+\dfrac{\partial\hat{u}_r}{\partial r}+\dfrac{\hat{u}_r}{r}+\dfrac{im}{r}\hat{u_\theta}\right)=0\mathrm{,}\\
& \hat{p}=\dfrac{\gamma -1}{\gamma}\left(\overline{\rho}\hat{T}+\overline{T}\hat{\rho}\right)\mathrm{.}
\end{empheq}
\label{eq:F_LEE}
\end{subequations}

The Fourier-transformed LEE \eqref{eq:F_LEE} can be written exclusively in terms of pressure, leading to the compressible Rayleigh equation \eqref{eq:Rayleigh}. The eigenfunctions $\hat{u}_i\left(r\right)$, $\hat{\rho}\left(r\right)$ and $\hat{T}\left(r\right)$ are recovered from the pressure

{\small
\begin{subequations}
\begin{align}
& \hat{u}_x\left(r\right)=-\frac{1}{\overline{\rho}\left(\overline{u}_xk-\omega\right)^2}\frac{\partial\overline{u}_x}{\partial r}\frac{\partial\hat{p}}{\partial r}-\frac{k}{\overline{\rho}\left(\overline{u}_xk-\omega\right)}\hat{p}\mathrm{,}\\
& \hat{u}_r\left(r\right)=-\frac{1}{\overline{\rho}\left(\overline{u}_xik-i\omega\right)}\frac{\partial\hat{p}}{\partial r}\mathrm{,}\\
& \hat{u}_\theta\left(r\right)=-\frac{m}{\overline{\rho}r\left(\overline{u}_xk-\omega\right)}\hat{p}\mathrm{,}\\
& \hat{\rho}\left(r\right)=-\frac{1}{\left(\overline{u}_xk-\omega\right)^2}\left(\frac{\partial^2\hat{p}}{\partial r^2}+\left(\frac{1}{r}-\frac{2k}{\overline{u}_xk-\omega}\frac{\partial\overline{u}_x}{\partial r}\right)\frac{\partial\hat{p}}{\partial r}-\left(k^2+\frac{m^2}{r^2}\right)\hat{p}\right)\mathrm{,}\\
& \hat{T}\left(r\right)=-\frac{\left(\gamma -1\right)\overline{T}}{\overline{\rho}\left(\overline{u}_xk-\omega\right)^2}\left(\frac{\partial^2\hat{p}}{\partial r^2}+\left(\frac{1}{r}-\frac{2k}{\overline{u}_xk-\omega}\frac{\partial\overline{u}_x}{\partial r}-\frac{1}{\overline{\rho}}\frac{\partial\overline{\rho}}{\partial r}+\frac{1}{\left(\gamma -1\right)\overline{T}}\frac{\partial\overline{T}}{\partial r}\right)\frac{\partial\hat{p}}{\partial r}-\left(k^2+\frac{m^2}{r^2}\right)\hat{p}\right)\mathrm{.}
\end{align}
\label{eq:eigenfunctions}
\end{subequations}}

\section{Energy norm derivation}
\label{sec:energy_norm_app}
Eigenfunctions are normalised to have zero phase for the pressure eigenfunction on the centreline and to have unitary energy norm. Following \cite{chu1965energy}, the energy norm is defined as,

\begin{equation}
E=\frac{1}{2}\int\limits_V\left(A u_i^*u_i + B\rho^*\rho+C T^*T\right)\mathrm{d}V\mathrm{,}
\end{equation}

\noindent where the $*$ indicates the complex conjugate and the constants $A$, $B$ and $C$ have to determined on the basis of the flow considered \citep{hanifi1996transient}. We assume a medium at rest, thus implying $\overline{u}_x=0$ and $\partial\overline{q}/\partial r=0$. Hence, the LEE reduce to the following,

\begin{subequations}
\begin{empheq}[left=\empheqlbrace]{align}
&\frac{\partial\rho}{\partial t}+\overline{\rho}\nabla\cdot\vec{u}=0\mathrm{,}\\
&\overline{\rho}\frac{\partial u_x}{\partial t}=-\frac{\partial p}{\partial x}\mathrm{,}\\
&\overline{\rho}\frac{\partial u_r}{\partial t}=-\frac{\partial p}{\partial r}\mathrm{,}\\
&\overline{\rho}\frac{\partial u_\theta}{\partial t}=-\frac{1}{r}\frac{\partial p}{\partial \theta}\mathrm{,}\\
&\frac{\partial T}{\partial t}+\left(\gamma -1\right)\overline{T}\nabla\cdot\vec{u}=0\mathrm{,}\\
&p=\frac{\gamma -1}{\gamma}\left(\overline{\rho}T+\overline{T}\rho\right)\mathrm{.}\label{eq:state}
\end{empheq}
\label{eq:LEE_E}
\end{subequations}

\noindent where we have removed the prime for the fluctuating part for notational simplicity. We calculate the time evolution of the energy norm,

\begin{equation}
\frac{\partial E}{\partial t}=\frac{\partial E}{\partial u_i}\frac{\partial u_i}{\partial t} + \frac{\partial E}{\partial \rho}\frac{\partial \rho}{\partial t} + \frac{\partial E}{\partial T}\frac{\partial T}{\partial t}\mathrm{,}
\end{equation}

\noindent which, exploiting \eqref{eq:LEE_E}, can be written as

\begin{equation}
\frac{\partial E}{\partial t}=-\frac{1}{2}\int\limits_V \left(A u_i\frac{1}{\overline{\rho}}\nabla p + B\,\rho\,\overline{\rho}\nabla\cdot\mathbf{u} + CT\left(\gamma -1\right)\overline{T}\nabla\cdot\mathbf{u}\right)\mathrm{d}V\mathrm{.}
\label{eq:de_dt}
\end{equation}

Expressing $\rho=\frac{\gamma}{\gamma -1}\frac{p}{\overline{T}}-\frac{\overline{\rho}}{\overline{T}}T$ from \eqref{eq:state} and applying integration by parts on the first term on the right hand-side of \eqref{eq:de_dt}, the time evolution of the energy norm becomes,

\begin{equation}
\frac{\partial E}{\partial t}=-\int\limits_{\partial V}u_ip\,\mathrm{d}S + \int\limits_V \left(\frac{A}{\overline{\rho}}p - B\overline{\rho}\left(\frac{\gamma}{\gamma -1}\frac{p}{\overline{T}}-\frac{\overline{\rho}}{\overline{T}}T\right) - CT\left(\gamma -1\right)\overline{T}\right)\nabla\cdot\mathbf{u}\,\mathrm{d}V\mathrm{,}
\end{equation}

\noindent where $\int\limits_{\partial V}u_ip\,\mathrm{d}S$ represents the acoustic power crossing the domain $V$. Hence, $\partial E/\partial t+\int\limits_{\partial V}u_ip\,\mathrm{d}S$ can be interpreted as the total energy variation, which we impose to be conservative thus leading to the following expression,

\begin{equation}
\int\limits_V \left(\frac{A}{\overline{\rho}} - B\frac{\overline{\rho}}{\overline{T}}\frac{\gamma}{\gamma -1}\right)p\nabla\cdot\mathbf{u}\,\mathrm{d}V + \int\limits_V \left(B\frac{\overline{\rho}^2}{\overline{T}} - C\left(\gamma -1\right)\overline{T}\right)T\nabla\cdot\mathbf{u}\,\mathrm{d}V = 0\mathrm{.}
\label{eq:de_dt_0}
\end{equation}

Equation \eqref{eq:de_dt_0} can be equal to zero if and only if

\begin{subequations}
\begin{empheq}[left=\empheqlbrace]{align}
&\frac{A}{\overline{\rho}} - B\frac{\overline{\rho}}{\overline{T}}\frac{\gamma}{\gamma -1}=0\mathrm{,}\\
&B\frac{\overline{\rho}^2}{\overline{T}} - C\left(\gamma -1\right)\overline{T}=0\mathrm{.}
\end{empheq}
\end{subequations}

By legitimately imposing $A=\overline{\rho}$, it is straightforward to calculate the values for the other two constants:

\begin{subequations}
\begin{empheq}[left=\empheqlbrace]{align}
&B=\frac{\gamma -1}{\gamma}\frac{\overline{T}}{\overline{\rho}}\mathrm{,}\\
&C=\frac{\overline{\rho}}{\gamma\overline{T}}\mathrm{.}
\end{empheq}
\end{subequations}

For an inviscid locally parallel flow with medium at rest, the energy norm is finally given by \eqref{eq:energy_norm}.

\section{Mean radial velocity downstream of the shock}
\label{sec:radial_mean}
The presence of the shock wave generates a mean pressure gradient along the radial direction downstream of the shock. This induces the appearance of a mean radial velocity. This radial velocity is here evaluated in the case of a finite thickness model in order to establish that it does not affect the calculation of the reflection coefficient. We consider the momentum equation along the radial direction \eqref{eq:momentum_r}. The conservation equation for the mean flow is,

\begin{equation}
\overline{\rho}\overline{u}_r\frac{\partial\overline{u}_r}{\partial r}=-\frac{\partial\overline{p}}{\partial r}\mathrm{,}
\end{equation}

\noindent which gives for the mean radial velocity the following expression,

\begin{equation}
\overline{u}_r=\sqrt{-2\int\limits_0^\infty\frac{1}{\overline{\rho}}\frac{\partial\overline{p}}{\partial r}\,\mathrm{d}r}\mathrm{.}
\end{equation}

\begin{figure}
\centering
\includegraphics[scale=0.25]{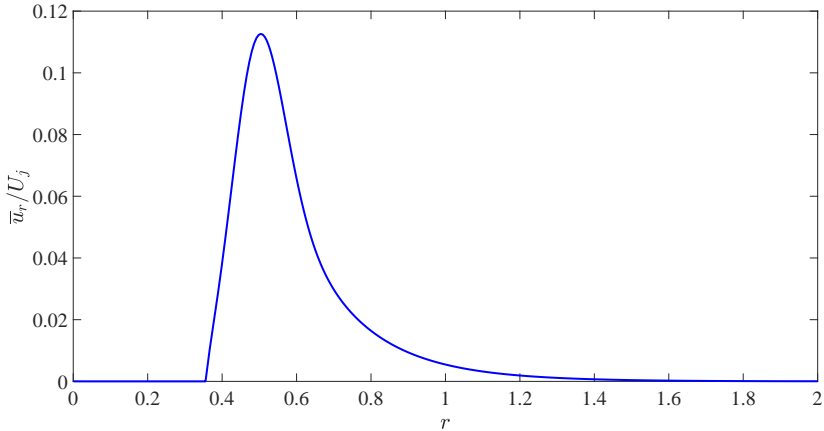}
\caption{Mean radial velocity profile along the radial direction downstream of the shock for $M_{j_1}$ and $M_{j_2}$ equal to 1.1 and 0.91, respectively.}
\label{fig:ur_mean}
\end{figure}

The radial derivative is computed using a centred finite-difference method. The radial profile of the mean radial velocity normalised by the jet Mach number downstream of the shock is reported in figure \ref{fig:ur_mean}. We note that $\overline{u}_r\neq 0$ in the shear-layer region with a peak for $r=0.5$. Nevertheless, since the maximum value is about 10\% of the jet velocity, we may assert that the jet is slowly diverging downstream of the shock, but this feature does not affect the evaluation of the reflection coefficient given that both the reflection and transmission mechanisms happen locally and are not influenced by the flow evolution far away from the shock.

\section*{Acknowledgements}
The authors acknowledge the financial support of EU and Nouvelle-Acquitaine region under the program CPER-FEDER. M.M. acknowledges the support of Centre National d'\'{E}tudes Spatiales (CNES) under a post-doctoral grant at the time of the writing of the paper. E.M. acknowledges the financial support of Commissariat à l'Energie Atomique, Centre d'Etudes Scientifiques et Techniques d'Aquitaine (CEA-CESTA) under a post-doctoral grant.

\section*{Declaration of interest}
The authors report no conflict of interest.

\bibliographystyle{jfm}
\bibliography{biblio}

\end{document}